\documentclass[%
 reprint,
 amsmath,amssymb,
 aps,
prb,
]{revtex4-1}

\usepackage{graphicx}% Include figure files
\usepackage{dcolumn}% Align table columns on decimal point
\usepackage{bm}% bold math
\usepackage{xcolor}

\begin{document}
\title{Chirality-dependent planar Hall effect in inhomogeneous Weyl semimetals}
\author{Suvendu Ghosh$^1$}\email{suvenduphys@iitkgp.ac.in} \author{Debabrata Sinha$^2$, Snehasish Nandy$^{1,3}$} \author{A. Taraphder$^{1,2}$}\email{arghya@phy.iitkgp.ac.in}

\affiliation{$^1$ Department of Physics, Indian Institute of Technology,
Kharagpur-721302, India \\
$^2$ Center for Theoretical Studies, Indian Institute of Technology,
Kharagpur-721302, India\\
$^3$ Department of Physics, University of Virginia, Charlottesville, VA 22904, USA}
%\date{\today}

\begin{abstract}
The planar Hall effect (PHE), the appearance of an in-plane transverse voltage in the presence of co-planar electric ($\mathbf{E}$) and magnetic ($\mathbf{B}$) fields, occurs in regular Weyl semimetals (WSMs) as one of the fundamental manifestations of chiral anomaly. A major
issue, therefore, is whether there are alternate route to PHE, without invoking chiral anomaly. We demonstrate that PHE exists in an inhomogeneous Weyl semimetal (IWSM) even in the absence of the aforesaid anomaly. Using semiclassical Boltzmann transport theory, we show that PHE appears in an IWSM due to the strain-induced chiral gauge potential, which couples to the Weyl fermions of opposite chirality with opposite sign. Our study shows a resultant phase shift in the current associated with opposite chirality Weyl nodes, which, remarkably, leads to a finite chirality-dependent planar Hall effect (CPHE) in the IWSMs. Interestingly, we show that a small tilt in the Weyl node can generate a pure CPHE even in the absence of an applied magnetic field. The CPHE has important implications in `chiralitytronics'. We also discuss the experimental feasibility of these novel effects of strain in type-I IWSMs.

\end{abstract}
\maketitle
%\section{Introduction}
\textbf{Introduction:-} Theoretical predictions and experimental discoveries of Weyl semimetals (WSMs) have led to an explosion of activities in the area of three-dimensional topological systems in recent years. Weyl semimetals appear as topologically nontrivial conductors where the spin-nondegenerate valence and conduction bands touch at isolated points in momentum space, the so-called ``Weyl nodes" \cite{Murakami07,Murak07,Wan11,Yang11,Burkov11,BB11,Xu11,Armitage18}. The Weyl nodes with opposite chirality appear in pairs in the absence of time-reversal symmetry (TRS) or inversion symmetry\cite{Wan11,Yang11,Burkov11,BB11,Xu11,Zyuzin12,
Zyuz12,Meng12}. Thus the chirality of Weyl fermions is an emergent property and ubiquitous in WSMs. The possibility to probe and manipulation of it, remain a challenge both in experiment as well as in theory.

Another intriguing property of a WSM is the chiral anomaly i.e., the anomalous non-conservation of chiral current in the presence of external fields parallel to each other. The appearances of negative magnetoresistance and PHE in WSMs are the indications of the chiral anomaly, which have been attracted intense experimental and theoretical interest\cite{Nielsen81,Nielsen83,Goswami15,Goswami13,Zhong15,Bell69,Aji12,Adler69,Burkov15,Nandy17,Nandy19,Sharma16,Burkov17,Huang15,Jia16,Li17,Wang16}. PHE in WSMs manifests itself when the applied current, magnetic field, and the induced transverse voltage all lie in the same plane, precisely the configuration where the conventional Hall effect vanishes \cite{Nandy17}. This is indeed in contrast to the chiral magnetic effect (CME) which refers to an electric current flowing along the direction of the applied magnetic field triggered by the chirality imbalance in the Weyl nodes. The difference between CME and PHE is that the CME current appears always along the direction of the applied magnetic field and without any electric field whereas the planar Hall current flows perpendicular to the applied in-plane electric field irrespective of the direction of the applied in-plane magnetic field. However, with the discovery of PHE, a natural question immediately arises whether it can be used as a chirality probe in a WSM. Unfortunately, chirality is ill-defined in this situation since the applied magnetic field can't differentiate the opposite chiral nodes. However, in the case of an inhomogeneous Weyl semimetal (IWSM), the pseudo-magnetic field couples opposite chiral Weyl fermions with an opposite signs~\cite{Behrends19,Liu13,Chernodub14,Grushin16,Pikulin16,Cortijo15,Cortijo16,Zhou-CPL13} which, paves the way toward the realization of chirality dependent PHE (CPHE).

The spatial or temporal variation of Weyl nodes separation $(b_0,\mathbf{b})$ generate chiral pseudomagnetic field ($\mathbf{B_5}=\mathbf{\nabla}\times\mathbf{b}$) and pseudoelectric field ($\mathbf{E_5}=-\mathbf{\nabla}b_0-\partial_t\mathbf{b}$)\cite{Grushin16,Pikulin16,Cortijo15}. \textcolor{black}{In the presence of both electromagnetic and pseudo-electromagnetic fields, the covariant anomaly relations can be written as \cite{Liu13,Grushin16,Pikulin16}}:%can be varied both spatially as well as temporally by means of lattice deformations
%due to the elastic deformations of the lattice 
%originating from a mechanical strain or inhomogeneous magnetization \cite{Liu13,Chernodub14,Grushin16,Pikulin16,Cortijo16}. In an inhomogeneous Weyl semimetal (IWSM) or strain-induced WSM, the space- and time-dependent $\mathbf{b}$ and/or time-dependent $b_0$ generate chiral pseudomagnetic field ($\mathbf{B_5}=\mathbf{\nabla}\times\mathbf{b}$) and pseudoelectric field ($\mathbf{E_5}=-\mathbf{\nabla}b_0-\partial_t\mathbf{b}$) which, unlike the real electromagnetic fields, couple with opposite signs to opposite chiralities~\cite{Liu13,Chernodub14,Grushin16,Pikulin16,Cortijo15,Zhou-CPL13}.

\begin{eqnarray}
\label{con-1}
\partial_t\rho_5+\mathbf{\nabla}\cdot\mathbf{j_5}&=&\frac{e^2}{2\pi^2\hbar^2}(\mathbf{E}\cdot\mathbf{B}+\mathbf{E_5}\cdot\mathbf{B_5})\\
\partial_t\rho+\mathbf{\nabla}\cdot\mathbf{j}&=&\frac{e^2}{2\pi^2\hbar^2}(\mathbf{E}\cdot\mathbf{B_5}+\mathbf{E_5}\cdot\mathbf{B}),
\label{con-2}
\end{eqnarray}
where $\rho$ and $\rho_5$ are the total electron and chiral density, respectively, and $\mathbf{j}$ and $\mathbf{j_5}$ are current densities associated to $\rho$ and $\rho_5$ respectively. $\rho_5=\rho_R-\rho_L$, represents the difference between the charge densities associated with the right- and left-handed Weyl points. Eq.~(\ref{con-1}) expresses the famous chiral anomaly represented by real fields $(\mathbf{E \cdot B})$ or pseudo-electromagnetic fields $(\mathbf{E_5 \cdot B_5})$.
% and refers to the generation of an electric current parallel to a magnetic field whenever pumping of charge happens from one Weyl point to the other due to an imbalance between the number of right- and left-handed Weyl fermions. 
Eq.~(\ref{con-2}) suggests that the total charge conservation is violated if either $\mathbf{E_5}\neq0$ or $\mathbf{B_5}\neq0$ in the presence of real external fields. For a TRS-broken but inversion symmetric IWSM having no temporal variation of $\mathbf{b}$, there will be no pseudo-electric field present and the violation of total charge conservation is apparently due to the term $\propto \mathbf{E}\cdot\mathbf{B_5}$. \textcolor{black}%The local non-conservation of charge corresponds to the pumping of charge between the bulk and the edge of the system~\cite{Grushin16,Pikulin16}. The set of anomaly relations (Eqs.\ref{con-1},\ref{con-2}) are the covariant anomaly in quantum field theory language. 
%{In order to satisfy the charge conservation locally, one should redefine electric four-current density with taken into consideration of topological Chern-Simons contribution\cite{Bardeen-NPB84}. The corrected physical currents are the consistent currents which satisfy the local charge conservation law\cite{Land-PRB14}.}
{Now, in order to conserve the electric charge locally as well as to get the corresponding {\it consistent currents\cite{Bardeen-NPB84}}, one should redefine electric four-current density taking into consideration the topological {\it Chern-Simons contribution}\cite{Land-PRB14}.}

%On the other hand, it is clear from the Eq.~(\ref{2}) that the total charge conservation is violated if either $\mathbf{E_5}\neq0$ or $\mathbf{B_5}\neq0$ in the presence of real external fields. This apparent charge-density non-conservation which results in pumping of charge between the bulk and the edge of the system gives a novel kind of anomaly in an IWSM~\cite{Pikulin16,Behrends19}. For a TRS-broken but inversion symmetric IWSM having no temporal variation of $\mathbf{b}$, there will be no pseudo-electric field present and the violation of total charge conservation is apparently due to the term $\propto \mathbf{E}\cdot\mathbf{B_5}$.

%Weyl semimetals exhibit several fascinating anomaly-related transport properties~\cite{Nielsen81,Nielsen83,Goswami15,Goswami13,Zhong15,Bell69,Aji12,Adler69,Burkov15,Nandy17,Nandy19,Sharma16,Burkov17,Huang15,Jia16,Li17,Wang16}. One such phenomenon is the occurrence of positive longitudinal magneto-conductivity (LMC) and the associated planar Hall effect (PHE) due to the proper chiral anomaly $\mathbf{E}\cdot\mathbf{B}$~\cite{Nandy17,Burkov17}. 

A TRS-breaking IWSM, where $\mathbf{E_5}=0$, presents a fascinating ground for the search of novel transport signatures due to the presence of the anomaly. The anomaly is exhibited through the enhancement of longitudinal magnetoconductivity (LMC) as $\sigma \sim \mathbf{B}^2_5$ due to the pseudo-magnetic field~\cite{Grushin16}. It is natural to query how this  anomaly is manifested in PHE. More importantly, whether PHE and CPHE can be realized in the absence of chiral anomaly i.e., solely due to the underlying pseudo-magnetic field.

Our study for a TRS broken but inversion symmetric type-I IWSM shows that PHE can exists due to the pseudo magnetic field and the related anomaly. The field $\mathbf{B_5}$ acting with an opposite sign at two opposite chiral nodes.  Thus, it brings a novel chirality-dependent PHE (CPHE), which is defined as the difference of contributions from the two nodes of different chiralities. In a nontilted IWSM, CPHE is finite when $\mathbf{B}$ and $\mathbf{B_5}$ are both presents and remarkably, a small tilt in Weyl nodes can lead to a finite CPHE even in the absence of real magnetic field. Thus CPHE presents a new scheme to manipulate the chirality index in a WSM. Recent experimental realizations~\cite{Peri19,Kamboj19,Jia19} of IWSMs add to the quest for these outcomes.

\textbf{Boltzmann formalism in an inhomogeneous Weyl Semimetal:-}
The momentum space Hamiltonian for a linearized tilted Weyl node can be expressed as 
\begin{eqnarray}
H_{\chi}=\hbar v_F(\chi \mathbf{k}\cdot \mathbf{\sigma}+\mathbf{\gamma}_{\chi}\cdot \mathbf{k}\sigma_0)-\mu,
\label{tilt-hamil}
\end{eqnarray}
where $v_F$ is the Fermi velocity, $\chi$ is the chirality associated with the Weyl node, $\mathbf{\sigma}$ represents the vector of Pauli matrices, $\sigma_0$ is the identity matrix, and $\mathbf{\gamma}_{\chi}$ is the tilt parameter along arbitary $\hat{k}$-direction. We choose that tilting is along the $k_x$-direction for the rest of our work without any loss of generality~\cite{Soluyanov15}. In this work, we consider both the cases of chiral and non-chiral tilt in case of time reversal broken WSM with two tilted type-I Weyl cones. In case of chiral tilt, two Weyl cones are tilted in the opposite direction exactly by the same amount, i.e., $\mathbf{\gamma_\chi}=\chi \gamma_x\hat{k}_x$. But non-chiral tilt simply means $\mathbf{\gamma_\chi}=\gamma_x\hat{k}_x$, i.e., the Weyl cones tilt in the same direction by the same amount. Here we restrict our discussion for type-I WSM ($\gamma_x<1$) where we always have the point-like Fermi surface at the Weyl node. However, we'll use the natural unit for rest of this work, that is,  $\hbar=c=1$.

One of the key concepts in developing the formalism for planar Hall effect in an IWSM is that, in presence of external magnetic field as well as static strain applied to these systems, chiral charges feel different effective fields given by $\mathbf{B_{\chi}}=\mathbf{B}+\chi\mathbf{B_5}$ according to their chiralities~\cite{Grushin16,Behrends19,Roy18}. Now, we briefly present the semi-classical formula for the planar Hall conductivity (PHC) and longitudinal magneto-conductivity (LMC). We have performed our calculations taking $B\sim 0-3T$, $B_5\sim 0-3T$ and considering temperature $T\ll(\sqrt{B},\sqrt{B_5})\ll\mu$. The Landau quantization can be neglected in these low magnetic field ($\mathbf{B}$,$\mathbf{B_5}$) regime. The transport properties of electrons can be understood using the famous phenomenological Boltzmann transport equation\cite{Ziman1}.
%\textcolor{red}{In order to derive the cosistent electric currents in presence of both electromagnetic and pseudoelectromagnetic fields, one should use the consistent kinetic theory. However, we show in the Supplemental Material~\cite{suppli1} that, for the planar hall configuration, the topological Chern-Simons contribution to the planar Hall conductivity and longitudinal conductivity vanishes.}

To calculate planar Hall effect in the presence of real as well as pseudomagnetic fields, we apply the electric field along the $x$-axis and the real magnetic field in the $xy$ plane i.e., $\mathbf{E}=E\hat{x}$ and $\mathbf{B}=B\cos\theta_b\hat{x}+B\sin\theta_b\hat{y}$. We apply the strain in such a way that the pseudomagnetic field is also lying on the $xy$-plane i.e., $\mathbf{B}_5=B_5\cos\theta_s\hat{x}+B_5\sin\theta_s\hat{y}$. Here, $\theta_b$ ($\theta_s$) is the angle between electric field and real (pseudomagnetic) field. In the presence of the effective magnetic field $\mathbf{B}^{\chi}=(\mathbf{B}+\chi \mathbf{B_5})$ and electric field $\mathbf{E}$, the semiclassical equations of motion are given in the Supplemental Material~\cite{suppli1}.
%In a typical Weyl semimetal, for the scattering time $\tau\sim10^{-13}s$ and cyclotron frequency $\omega_c\sim(eB/0.11m_ec)$, $\omega_c\tau\sim0.3<1$. Then  and transport properties of electrons can be understood, in the presence of impurity scattering and spatially uniform fields, using the semiclassical Boltzmann kinetic equation
Considering spatially uniform field in steady state condition, retaining only the contributions due to linear response, the general expression of the longitudinal conductivity and planar Hall conductivity associated with each Weyl node with chirality $\chi$ are calculated, respectively, as
\begin{eqnarray}
\label{e8}
\sigma^{\chi}_{xx}&=&-\frac{e^2\tau}{(2\pi)^3}\int d^3k D^{\chi}_{\mathbf{k}}[v_x^2+2ev_{x}(\mathbf{B}_\chi\cdot \hat{x})(\mathbf{v}_{k}\cdot\mathbf{\Omega}^\chi_{k})\nonumber\\&&+e^2(\mathbf{B}_\chi\cdot \hat{x})^2(\mathbf{v}_{k}\cdot\mathbf{\Omega}^\chi_{k})^2]\left({\partial f_{eq}\over \partial \epsilon_{\mathbf{k}}}\right)\\
\sigma^{\chi}_{yx}&=&-\frac{e^2\tau}{(2\pi)^3}\int d^3k D^{\chi}_{\mathbf{k}}[v_{x}v_{y}+ev_{y}(\mathbf{B}_\chi\cdot \hat{x})(\mathbf{v}_{k}\cdot\mathbf{\Omega}^\chi_{k})\nonumber\\&&+ev_{x}(\mathbf{B}_\chi\cdot \hat{y})(\mathbf{v}_{k}\cdot\mathbf{\Omega}^\chi_{k})\nonumber\\&&+e^2(\mathbf{B}_\chi\cdot \hat{x})(\mathbf{B}_\chi\cdot \hat{y})(\mathbf{v}_{k}\cdot\mathbf{\Omega}^\chi_{k})^2]\left({\partial f_{eq}\over \partial \epsilon_{\mathbf{k}}}\right)
\label{e9}
\end{eqnarray}
where $\mathbf{v}_{\mathbf{k}}=\frac{\partial\epsilon_\mathbf{k}}{\partial\mathbf{k}_{\chi}}$ is the group velocity and $D^{\chi}_{\mathbf{k}}=[1+e(\mathbf{B}_{\chi}\cdot\mathbf{\Omega}_{\mathbf{k}}^{\chi})]^{-1}$ depicts the modification of the phase space volume in the simultaneous presence of magnetic field and Berry curvature $\mathbf{\Omega}_{\mathbf{k}}^{\chi}$~\cite{Duval6,Xiao6,Xiao10}. \textcolor{black}{It is noteworthy that the topological {\it Chern-Simons current} does not contribute to both the planar Hall and longitudinal electric currents in our setup~\cite{suppli1}.} The higher order corrections due to the inhomogeneous Fermi velocity is neglected in this work~\cite{Sundaram99,Yang15,Jiang15}.
From Eqs.(\ref{e8}, \ref{e9}), We now define the total conductivity $\sigma^T_{ij}$ and chirality dependent conductivity $\sigma^C_{ij}$ as follows:

\begin{eqnarray}
\label{def_T}
\sigma^{T}_{ij}=\sigma^{\chi=+1}_{ij}+\sigma^{\chi=-1}_{ij}\\
\sigma^{C}_{ij}=\sigma^{\chi=+1}_{ij}-\sigma^{\chi=-1}_{ij}
\label{def_C}
\end{eqnarray}

\textbf{Conductivities in an inhomogeneous Weyl Semimetal:}
Following the formalism described in the previous section, we have analytically calculated the total LMC and PHC in case of non-tilted as well as tilted type-I IWSMs. The expressions for total LMC ($\sigma_{xx}^T$), and PHC ($\sigma_{yx}^T$) for nontilted case are given by~\cite{suppli1}:
\begin{eqnarray}
\label{r1}
\sigma_{xx}^T&=\frac{e^2\mu^2\tau}{3\pi^2v_F}+\frac{e^4v^3_F\tau}{60\pi^2\mu^2}B^2+\frac{e^4v^3_F\tau}{60\pi^2\mu^2}B^{2}_{5}\nonumber\\
&+\frac{7e^4v^3_F\tau}{60\pi^2\mu^2}(B^2\cos^2\theta_b+B^{2}_{5}\cos^2\theta_s),\\
\sigma_{yx}^T&=\frac{7e^4v^3_F\tau}{60\pi^2\mu^2}(B^2\sin2\theta_b+B^{2}_{5}\sin2\theta_s)
\label{r2}
\end{eqnarray}
The chirality-dependent longitudinal magneto-conductivity and planar Hall conductivity (CPHC) in case of non-tilted IWSMs are calculated as~\cite{suppli1}:
\begin{eqnarray}
\label{r3}
\sigma_{xx}^{C}&=&BB_5\frac{e^4v^3_F\tau}{15\pi^2\mu^2}(4\cos\theta_b\cos\theta_s+\frac{1}{2}\sin\theta_b\sin\theta_s),\\
\sigma_{yx}^{C}&=&\frac{7e^4v^3_F\tau}{60\pi^2\mu^2}BB_5\sin(\theta_b+\theta_s).
\label{r4}
\end{eqnarray}
The conductivities $\sigma^{\chi}_{ij}$ of two opposite chirality sectors in Eqs.(\ref{e8},\ref{e9}) in nontilted case, are related by: $\sigma^{+}_{ij}(\mathbf{B})=\sigma^{-}_{ij}(-\mathbf{B})$ and $\sigma^{+}_{ij}(\mathbf{B_5})=\sigma^{-}_{ij}(-\mathbf{B_5})$. Thus, $\sigma^T_{ij}$ and $\sigma^C_{ij}$ are respectively even and odd functions of $\mathbf{B}$ or $\mathbf{B_5}$. This implies $\sigma^{T(C)}_{xx}$ and $\sigma^{T(C)}_{yx}$  both have a quadratic (linear) dependency on $\mathbf{B}$ or $\mathbf{B_5}$. In a normal WSM (where $B_5=0$) both $\Delta\sigma_{xx}^T=(\sigma_{xx}^T(\mathbf{B})-\sigma_{xx}^T(B=0))$ and $\sigma_{yx}^T$ varies respectively as $B^2\cos^2\theta_b$ and $B^2\sin 2\theta_b$ with respect to applied magnetic field. The strain manifests itself by generating $B_5$ and enhances the LMC as well as the PHC, which are illustrated in Eqs.(\ref{r1},\ref{r2}). This enhancement of conductivities is attributed to the chiral pseudo-magnetic effect\cite{Grushin16}. Now, remarkably, if we set $B=0$, $\sigma_{yx}^T$ as well as $\sigma_{xx}^T$ still remain finite. In that case $\Delta\sigma_{xx}^T$ and $\sigma_{yx}^T$ follows $B_5^2\cos^2\theta_s$ and $B_5^2\sin 2\theta_s$ dependence with pseudo-magnetic field. This result has a profound significance. As $B=0$, the familiar chiral anomaly $\sim\mathbf{E}\cdot\mathbf{B}$ is absent here. Consequently, the \textcolor{black}{covariant anomaly} equations simplify to: $\partial_t \rho+\mathbf{\nabla}\cdot \mathbf{j}\sim \mathbf{E}\cdot \mathbf{B_5}$ and $\partial_t \rho_5+\mathbf{\nabla}\cdot \mathbf{j_5}\sim 0$. \textcolor{black}{Furthermore, with $\mathbf{B}=0$, the topological charge density ($\rho_{CS}$) vanishes  and the current density ($\mathbf{j}_{CS}$) is: $\mathbf{j}_{CS}\sim (\mathbf{b}\times \mathbf{E})$, which is responsible for the strain-induced anomalous Hall effect in a WSM (see Supplemental Material~\cite{suppli1}). Thus, PHE may find its existence solely as a strain induced effect in IWSMs.} From Eqs.(\ref{r3},\ref{r4}) it is clear that  both the chiral dependent conductivities (CDCs) $\sigma^C_{xx}$ and $\sigma^C_{yx}$ vanish in absence of either $\mathbf{B}$ or $\mathbf{B}_5$. It is also evident that, CDCs are tunable by the angle $\theta_b$ for any strain configuration.
%The underlying physics can be understood as follows. The chirality dependent chemical potential $\mu_\chi$ in presence of $\mathbf{E}\cdot \mathbf{B}_5$ term is given as\cite{Grushin16},
%\begin{eqnarray}
%\mu_\chi=\big[\mu^3-\frac{3}{2}\chi v^3_F e^2\tau(\mathbf{E}\cdot\mathbf{B}+\chi\mathbf{E}\cdot\mathbf{B}_5)\big]^{\frac{1}{3}}
%\end{eqnarray}
%A TRS broken Weyl semimetal in equilibrium i.e., in absence of external field, has $\mu_+=\mu_-=\mu$. The change in absolute value of the chemical potential and hence the conductivity, has different values in the two nodes if both of $\mathbf{B}$ and $\mathbf{B}_5$ are present. %We have noticed that chirality-dependent conductivity (CDC) has linear dependence on $\mathbf{B}$ and $\mathbf{B}_5$, contrary to the quadratic dependence of total conductivity. 
%From Eqs.(\ref{r3},\ref{r4}) it is evident that, CDC are controllable   by $\theta_b$ for a specific strain configuration.
%The underlying physics can be understood as follows. The chirality dependent chemical potential $\mu_\chi$ in presence of $\mathbf{E}\cdot \mathbf{B}_5$ term is given as\cite{Grushin16},
%\begin{eqnarray}
%\mu_\chi=\big[\mu^3-\frac{3}{2}\chi v^3_F e^2\tau(\mathbf{E}\cdot\mathbf{B}+\chi\mathbf{E}\cdot\mathbf{B}_5)\big]^{\frac{1}{3}}
%\end{eqnarray}
%A TRS broken Weyl semimetal in equilibrium i.e., in absence of external field, has $\mu_+=\mu_-=\mu$. The change in absolute value of the chemical potential has different values in the two nodes iff both of $\mathbf{B}$ and $\mathbf{B}_5$ are present.\\

\begin{figure}
\includegraphics[width=1.65in]{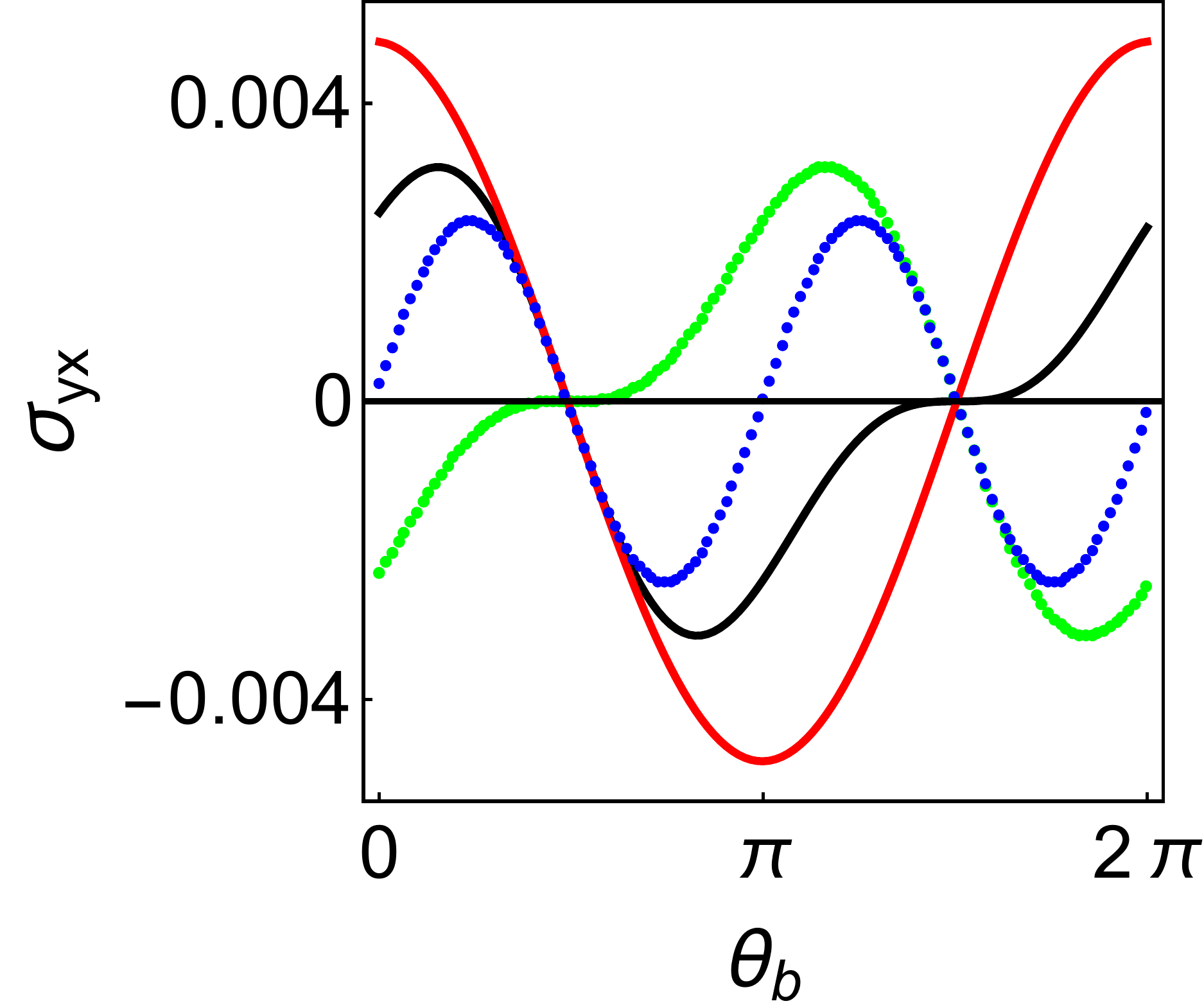}
\includegraphics[width=1.65in]{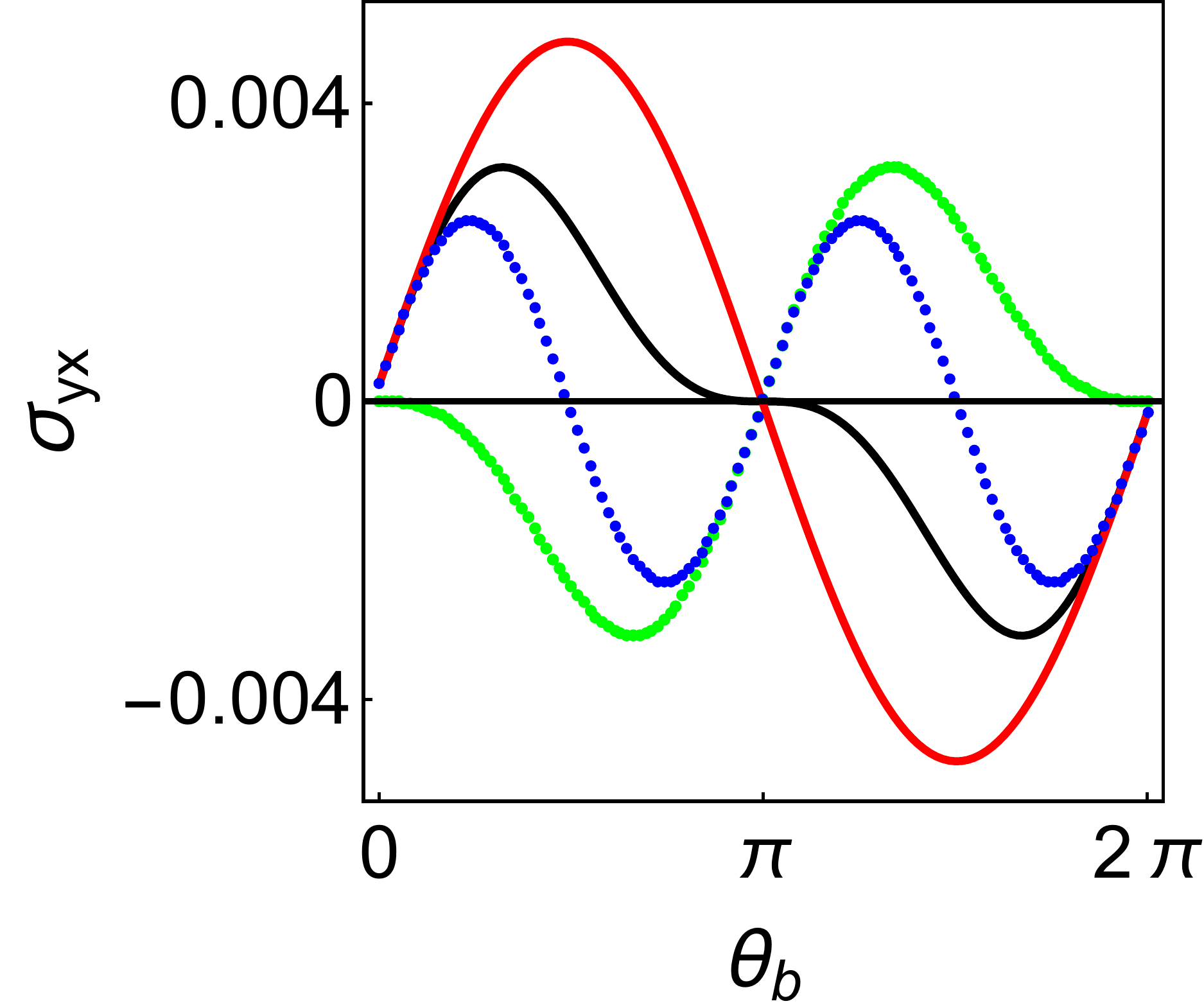}
\caption{Non-tilted IWSM. The angular ($\theta_b$) dependence of $\sigma^+_{yx}$ (black solid line), $\sigma^-_{yx}$ (green line), $\sigma^C_{yx}$ (red solid line) and $\sigma^T_{yx}$ (blue dotted line). We fix $\theta_s=\pi/2$ and  $\theta_s=0$ in left and right panel respectively. The other parameters are $B=B_5=0.5$T, $\mu=20$meV. All the conductivities are normalized by the corresponding total longitudinal conductivity without effective magnetic field $\sigma^T_{xx}(B=B_5=0)$. See text for details.}
\label{nont-fig1}
\end{figure}

%A finite chirality imbalance in the absolute value of chemical potential change between the nodes of opposite chirality causes $\sigma^+_{ij}\neq \sigma^{-}_{ij}$. This leads to a finite CDC which is one of the salient features in our present study. 

In absence of $\mathbf{B}_5$, the conductivities  $\sigma^{+}_{ij}$ and $\sigma^{-}_{ij}$ of two chiralities are identical. The total conductivity is twice of $\sigma^{+}_{ij}$ or $\sigma^{-}_{ij}$. A finite $\mathbf{B}_5$ shifts the phase oppositely of two opposite chiral nodes which leads a finite chiral dependent conductivity. We present our numerical results of angular dependence $\theta_b$ of $\sigma_{yx}$ in Fig.(\ref{nont-fig1}) for $\theta_s=\pi/2$ (in the left panel) i.e., $\mathbf{B}_5 \perp \mathbf{E}$  and $\theta_s=0$ (in the right panel) i.e., $\mathbf{B}_5 \parallel \mathbf{E}$. The corresponding analytical form of PHC of two chiralities are given: $\sigma^{\pm}_{yx}(\theta_s=\pi/2)=C_1 B^2\sin2\theta_b \pm C_2 B B_5\cos \theta_b$ and $\sigma^{\pm}_{yx}(\theta_s=0)=C_1 B^2\sin2\theta_b \pm C_2 B B_5\sin \theta_b$, with $C_1$ and $C_2$ are constants independent of $B$ and $B_5$. The black and green lines in Fig.(\ref{nont-fig1}) are displayed for $\sigma^+_{yx}$ and $\sigma^-_{yx}$, respectively. It is evident that PHC of two opposite chiralities are different, i.e., $\sigma^+_{yx} (\theta_b)\neq \sigma^-_{yx}(\theta_b)$, except at some specific values of $\theta_b$ where they both become zero. PHC of each chirality $\sigma^\chi_{yx}(\theta_b)$ vanishes at $\theta_b=(2n+1)\pi/2$ and $\theta_b=n\pi$ respectively, for $\theta_s=\pi/2$ and $\theta_s=0$. Also, PHC of opposite chiralities are related by: $\sigma^-_{yx}(\theta_b)=\sigma^+_{yx}(\theta_b \pm \pi)$. This implies that CPHC and total PHC are satisfying: $\sigma^C_{yx}(\theta_b \pm \pi)=-\sigma^C_{yx}(\theta_b)$ and $\sigma^T_{yx}(\theta_b \pm \pi)=\sigma^T_{yx}(\theta_b)$, respectively. Thus the important difference between CPHC and total PHC is that the former has $2\pi$ periodicity whereas later has $\pi$ periodicity with $\theta_b$. This holds true also for tilted IWSMs, discussed later. The solid red line in each panel of Fig.~\ref{nont-fig1} displays the results for CPHC and blue dotted line displays the total PHC, respectively. Moreover, $\sigma^C_{yx}(\theta_b)$ is an even function (i.e., $\sigma^C_{yx}(\theta_b)=\sigma^C_{yx}(-\theta_b)$) for $\theta_s=\pi/2$ and odd function (i.e., $\sigma^C_{yx}(\theta_b)=-\sigma^C_{yx}(-\theta_b)$) for $\theta_s=0$, respectively. Correspondingly, this leads to a finite CPHC at $\theta_b=n\pi$ (in left panel of Fig.~\ref{nont-fig1}) and $\theta_b=(2n+1)\pi/2$ (in right panel of Fig.~\ref{nont-fig1}), where the total PHC remains zero. This is one of the striking results here: the situation is akin to the pure valley current in valleytronics\cite{Rycerz07} and we call this effect pure CPHE. The important point we have noticed is that a pure CPHE occurs only for $\theta_s=\pi/2$ or $\theta_s=0$ (see Eq.(\ref{r2}) and Eq.(\ref{r4})). 
%\textcolor{blue}{The geometry to produce $\mathbf{B}_5$ field along $\theta_s=0$ and $\theta_s=\pi/2$ are discussed in the Supplemental Material~\cite{suppli1}.}
%\begin{center}
\begin{figure}
\includegraphics[width=1.65in]{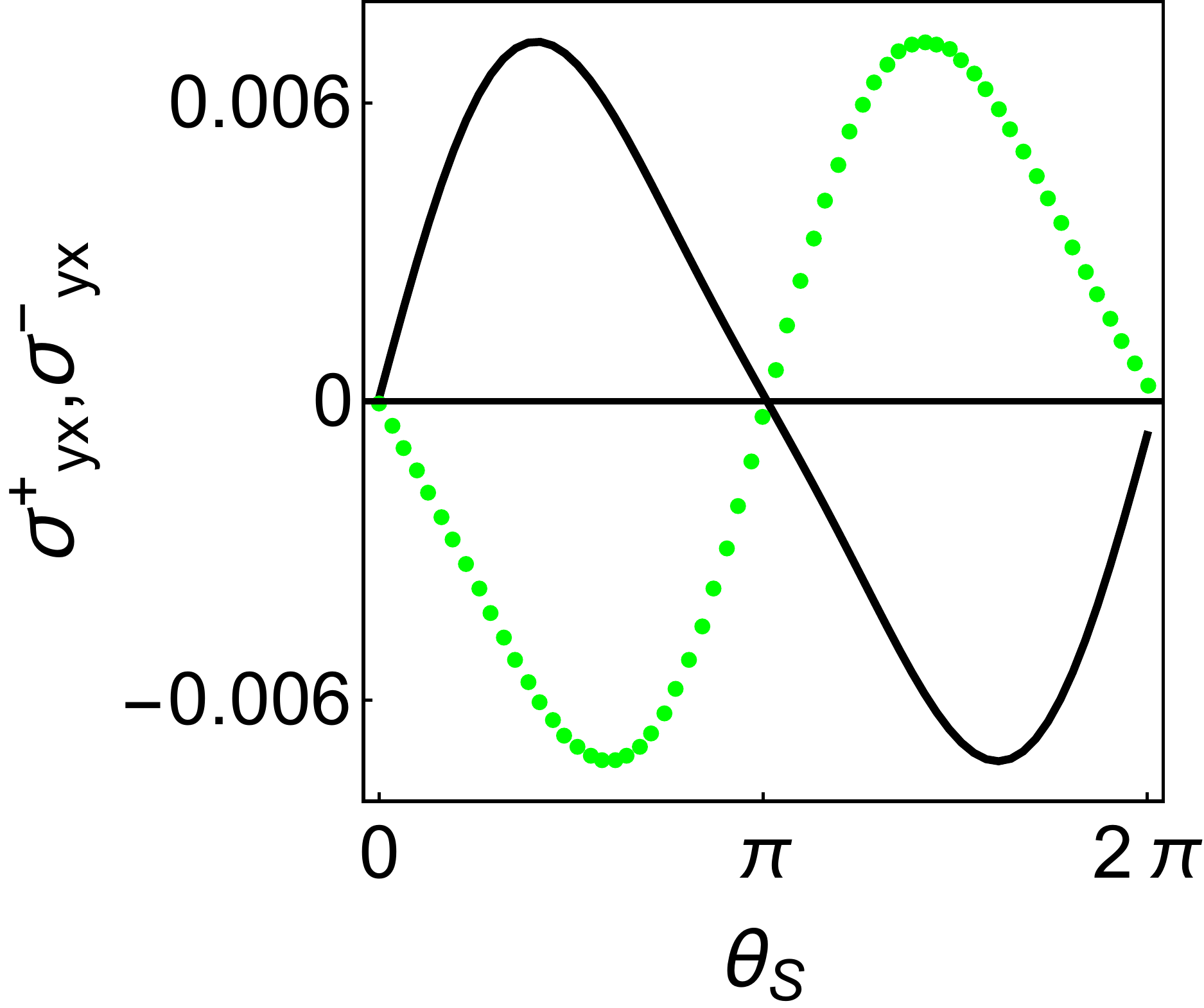}
\includegraphics[width=1.65in]{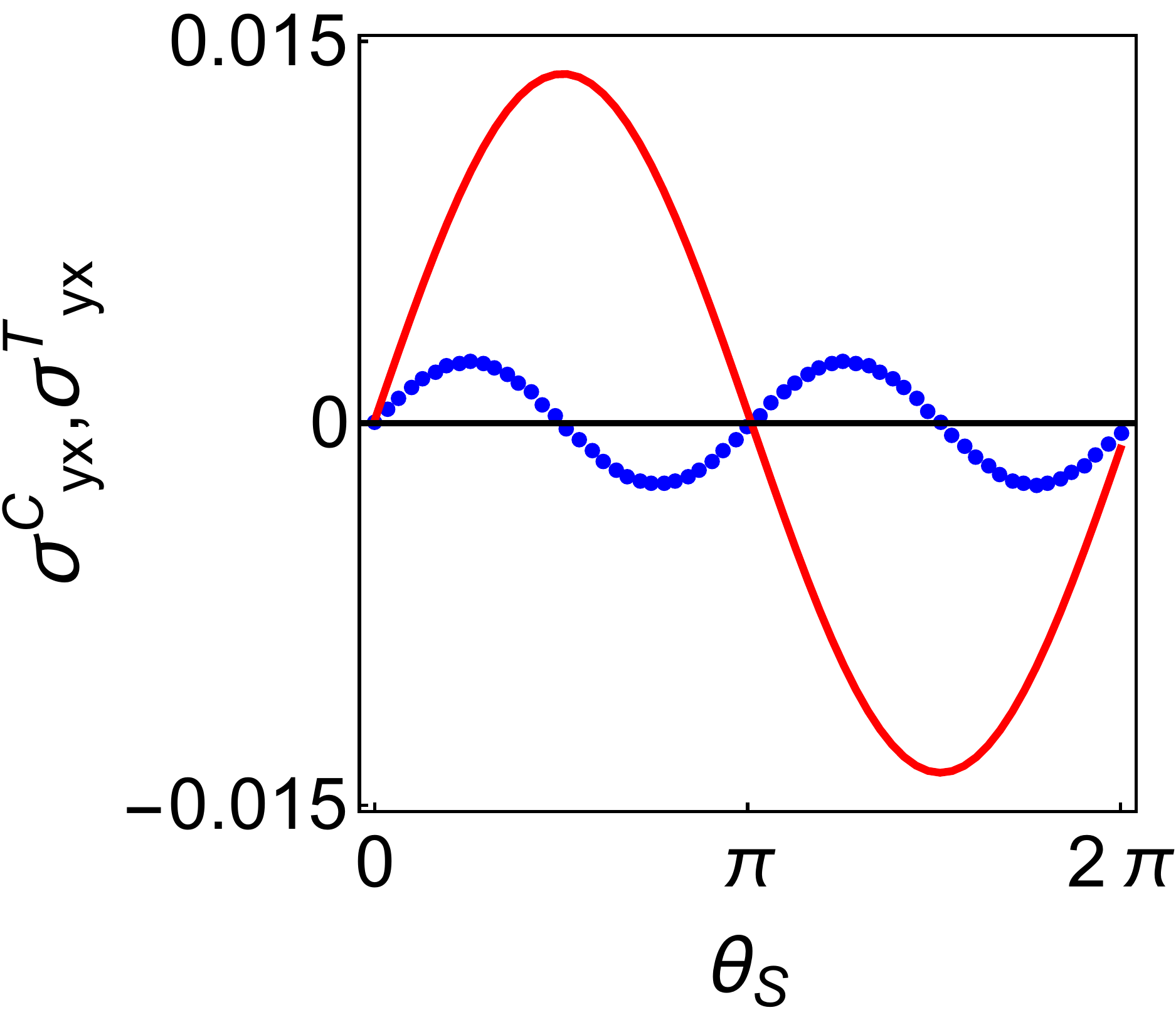}
\caption{Tilted IWSM (chiral-tilt). Left panel shows the planar Hall conductivity $\sigma^+_{yx}$ (black solid line) and $\sigma^-_{yx}$ (green dotted line) from each nodes of a tilted Weyl semimetal. Right panel shows Chirality dependent planar Hall conductivity $\sigma^C_{yx}$ (red solid line) and total planar Hall conductivity $\sigma^T_{yx}$ (blue dotted line). We set the parameters are $B=0$, $B_5=0.5$T and $\theta_b=0$. All the plots are for $\gamma_x=0.2$ and the conductivities are normalized by the corresponding total longitudinal conductivity without effective magnetic field $\sigma^T_{xx}(B=B_5=0)$.}
\label{fig1}
\end{figure}

Now we move to our discussions on the tilted type-I IWSM. Here, our analytical calculations reveal the total LMC and PHC for an IWSM with chiral tilt, as follows~\cite{suppli1}:
\begin{eqnarray}
\sigma_{xx}^T&=\frac{e^2\mu^2\tau}{\pi^2v_F}(\frac{1}{3}+2\gamma_x^4)+\frac{2e^4v^3_F\tau}{15\pi^2\mu^2}(B^2\cos^2\theta_b+B_5^2\cos^2\theta_s)\nonumber\\
&+\frac{e^4v^3_F\tau}{60\pi^2\mu^2}(1+7\gamma_x^2)(B^2\sin^2\theta_b+B_5^2\sin^2\theta_s)\nonumber\\
&+\frac{e^3v_F\tau}{6\pi^2}(\frac{23}{5}\gamma_x+\gamma_x^3)B\cos\theta_b,
\label{r5}
\end{eqnarray}
\begin{eqnarray}
\sigma_{yx}^T&=\frac{e^4v^3_F\tau}{40\pi^2\mu^2}(\frac{7}{3}+\gamma_x^2)(B^2\sin 2\theta_b+B_5^2\sin 2\theta_s)\nonumber\\&+\frac{e^3 v_{F}\tau}{15\pi^2}(\frac{11}{2}\gamma_x+\gamma_x^3)B\sin\theta_b
\label{r6}
\end{eqnarray}

\begin{center}
\begin{figure}
\includegraphics[width=1.69in]{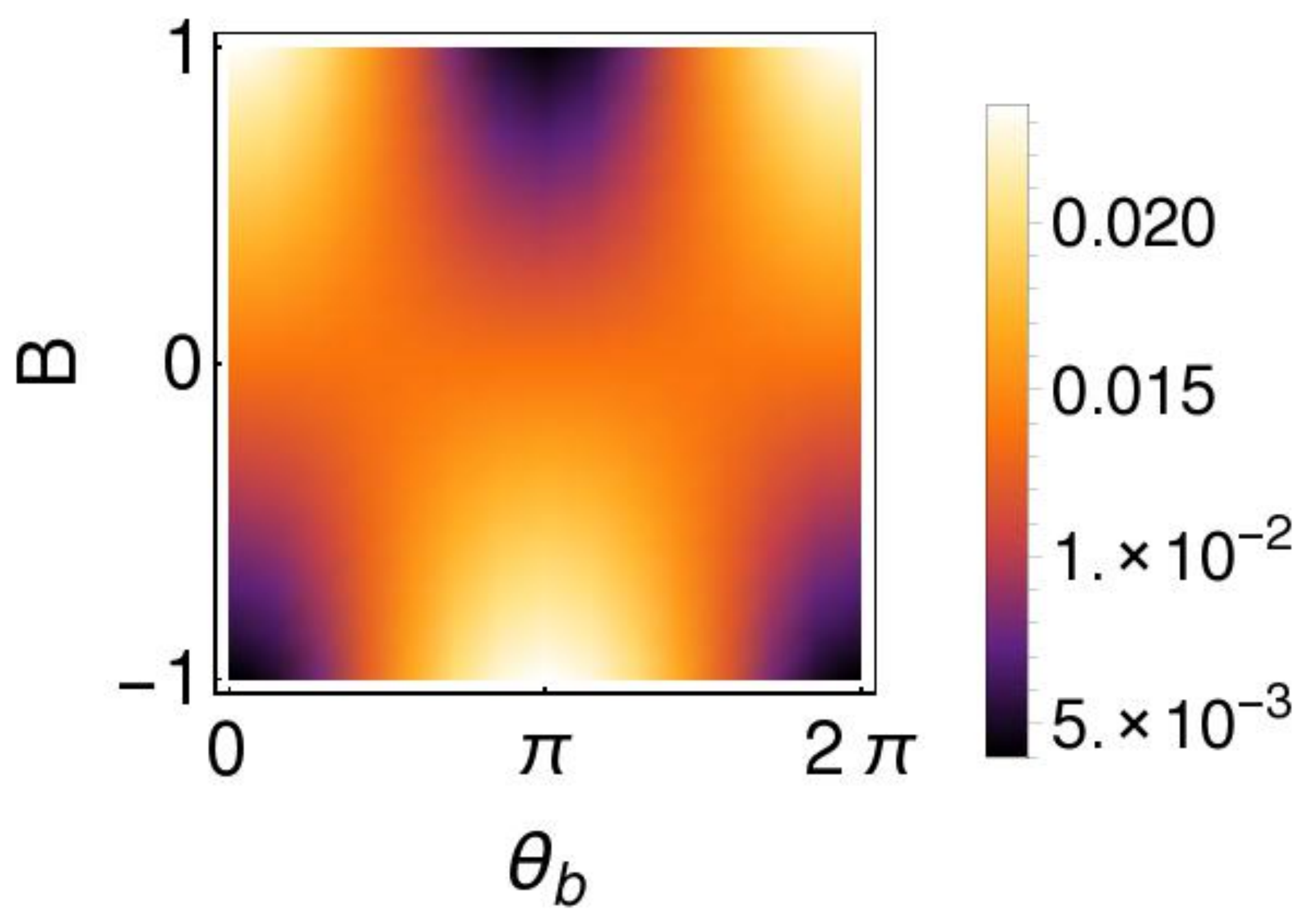}
\includegraphics[width=1.65in]{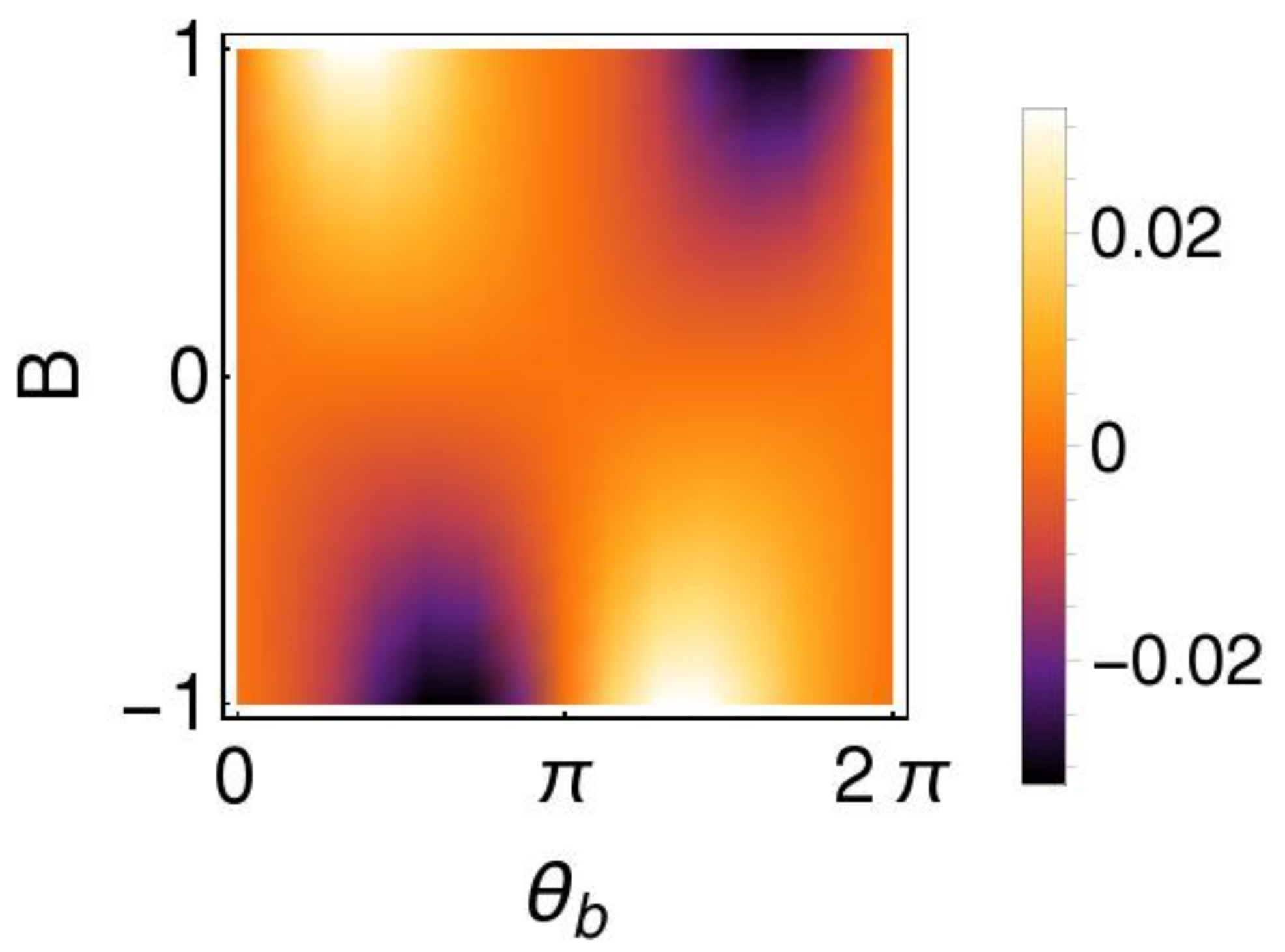}
\caption{Density plot of $\sigma^C_{yx}$ (left panel )and density plot of $\sigma^T_{yx}$ (right panel) with $B$ and $\theta_b$. We fix $B_5=0.5$T and $\theta_s=\pi/2$. All the plots are for $\gamma_x=0.2$ and the conductivities are normalized by the corresponding total longitudinal conductivity without effective magnetic field $\sigma^T_{xx}(B=B_5=0)$.}
\label{fig1-1}
\end{figure}
\end{center}

The total conductivities in Eqs.(\ref{r5},\ref{r6}) include quadratic variation of $\mathbf{B}$ and $\mathbf{B_5}$, along with a linear term in $\mathbf{B}$. In case of non-chiral tilt, the expressions for total LMC and PHC can be obtained by interchanging $B$ and $B_5$, $\theta_b$ and $\theta_s$ in Eqs. (\ref{r5}) and (\ref{r6}) respectively~\cite{suppli1}. So, both LMC and PHC are enhanced due to the chiral pseudomagnetic effect in tilted IWSM. Remarkably, for chiral (non-chiral) tilt, they exist even in absence of magnetic (pseudo-magnetic) field. This astonishing outcome is owing to the interplay between tilting term and $\mathbf{E}\cdot\mathbf{B_5}$ anomaly in IWSMs. The linear term in the conductivities contains odd power in $\gamma_x$. It is independent of $\mu$ since, the tilting causes a finite density of states near Weyl nodes. Thus the total conductivities can be tunable by the angular variation of fields at the nodes. It is clear that, $\Delta\sigma_{xx}^T$ and $\sigma_{yx}^T$, in the case of chiral tilt, respectively show $B_5^2\cos^2\theta_s$ and $B_5^2\sin2\theta_s$ dependencies whereas vary as $B_5\cos\theta_s$ and $B_5\sin\theta_s$ for non-chiral tilt respectively. Interestingly, we find that $\Delta\sigma_{xx}^T$ and $\sigma_{yx}^T$ in a non-chiral tilted IWSM, behave similarly with external magnetic field compare to the non-tilted case~\cite{suppli1}.

%$\Delta\sigma_{xx}^T=(\sigma_{xx}^T(\mathbf{B})-\sigma_{xx}^T(B=0))$ and $\sigma_{yx}^T$ for a normal WSM (where $B_5=0$) varies respectively as $B^2\cos^2\theta_b$ and $B^2\sin 2\theta_b$ with respect to applied magnetic field
%It is evident from Eqs.(\ref{r5},\ref{r6}) that the PHC as well as LMC are enhanced due to the chiral pseudomagnetic effect in presence of $\mathbf{B}_5$ and exist even when external magnetic field is switched off ($B=0$). This remarkable effect is attributed solely to the $\mathbf{E}\cdot\mathbf{B_5}$-type anomaly present in an IWSM. To find the exact nature of the magnetic field and angular dependencies of PHC and LMC, we have performed detailed numerical calculations. 

%It is clear that, $\Delta\sigma_{xx}^T$ and $\sigma_{yx}^T$, in the case of chiral tilt, respectively show $B_5^2\cos^2\theta_s$ and $B_5^2\sin2\theta_s$ dependencies whereas vary as $B_5\cos\theta_s$ and $B_5\sin\theta_s$ for non-chiral tilt respectively. Interestingly, we find that $\Delta\sigma_{xx}^T$ and $\sigma_{yx}^T$ in a non-chiral tilted IWSM, behave similarly with external magnetic field compare to the non-tilted case~\cite{suppli1}.

The analytical form of chirality-dependent longitudinal magneto-conductivity (CLMC) and planar Hall conductivity (CPHC) in case of chiral-tilted type-I IWSMs are calculated analytically as~\cite{suppli1}:
\begin{eqnarray}
\label{r9}
\sigma_{xx}^{C}&=&\frac{4e^4v^3_F\tau}{15\pi^2\mu^2}BB_5\cos\theta_b\cos\theta_s\nonumber\\&&+\frac{e^3v_F\tau}{6\pi^2}(\frac{23}{5}\gamma_x+\gamma_x^3)B_5\cos\theta_s\nonumber\\
&&+\frac{4e^4v^3_F\tau}{30\pi^2\mu^2}(1+7\gamma_x^2)BB_5\sin\theta_b\sin\theta_s\\
\sigma_{yx}^{C}&=&\frac{e^4v^3_F\tau}{20\pi^2\mu^2}(\frac{7}{3}+\gamma_x^2)BB_5\sin(\theta_b+\theta_s)\nonumber\\&&+\frac{e^3v_F\tau}{15\pi^2}(\frac{11}{2}\gamma_x+\gamma_x^3)B_5\sin\theta_s .
\label{r10}
\end{eqnarray}

%\begin{center}
%\begin{figure}
%\includegraphics[width=1.65in]{phw-fig3.pdf}
%\includegraphics[width=1.65in]{phw-fig4.pdf}
%\caption{(a) Longitudinal conductivity $\sigma^+_{xx}$ (black solid line) and $\sigma^-_{xx}$ (green dotted line) from each nodes of a tilted Weyl semimetal. (b) Plot of Chirality dependent longitudinal conductivity $\sigma^C_{xx}$ (red solid line) and total longitudinal conductivity $\Delta\sigma^T_{xx}$ (blue dotted line). We set the paramaters are $B=0$, $B_5=0.5$T and $\gamma_x=0.2$. All the conductivities are normalized by the corresponding total longitudinal conductivity without effective magnetic field $\sigma^T_{xx}(B=B_5=0)$. }
%\label{fig2}
%\end{figure}
%\end{center}
%\begin{center}
%\begin{figure}
%\includegraphics[width=1.3in]{phwt-fig5.pdf}
%\includegraphics[width=1.3in]{phwt-fig4.pdf}
%\caption{(a) Plot of $\Delta\sigma^T_{xx}(\theta_s=0)$ and $\sigma^T_{yx}(\theta_s=\pi/4)$ with $B_5$.(b) Linear dependency with $B_5$ of chiral dependent conductivity. $\sigma^C_{xx}(\theta_s=0)$ and $\sigma^C_{yx}(\theta_s=\pi/2)$ are shown by red and blue line respectively. Here, we fix $B=0$, $\gamma_x=0.2$.}
%\label{fig3}
%\end{figure}
%\end{center}
%The dependence of $\sigma_{ij}$ with angle $\theta_s$ and field $B_5$ changes for a finite tilt. 
%\textcolor{red}{We find a novel effect of $B_5$; CDC has finite value even in absence of real magnetic field $B$.}
It is evident from Eqs.(\ref{r9},\ref{r10}) that, in case of chiral tilt even in the absence of $\mathbf{B}$, $\sigma^{C}_{ij}$ is finite, becomes independent of $\mu$ and depends only on the odd powers of $\gamma_x$. This is a remarkable feature shown by a TRS broken tilted type-I IWSM. The expressions for chirality-dependent conductivities in case of non-chiral tilt can be obtained by interchanging $B$ and $B_5$, $\theta_b$ and $\theta_s$ in Eqs. (\ref{r9}) and (\ref{r10}) respectively~\cite{suppli1}. Thus CDC are also tunable at the Weyl nodes. Interestingly, $\sigma^{C}_{ij}$ vanishes for an IWSM with non-chiral tilt in absence of $\mathbf{B}$. 
%as can be seen from Eqs.(\ref{r11},\ref{r12}).
A finite CDC in case of chiral tilt implies a difference of $\sigma^\chi_{ij}$ between the two nodes i.e., $\sigma^+_{ij}(\theta_b,\theta_s)\neq \sigma^-_{ij}(\theta_b,\theta_s)$. The angular $\theta_s$ dependence of $\sigma^{\chi}_{yx}$ of a tilted IWSM with $B=0$ is shown in the left panel of Fig.~\ref{fig1}. PHC in the nodes of different chirality are related: $\sigma^-_{yx}(\theta_s)=\sigma^+_{yx}(\theta_s\pm\pi)$. This is due to the fact that the linear term of PHC depends only on odd powers of $\gamma_x$ and thus has opposite sign in the two chiralities sector. We show the variation of $\sigma^T_{yx}(\mathbf{B}=0)$ and $\sigma^C_{yx}(\mathbf{B}=0)$ with $\theta_s$ for a finite $\gamma_x$, in the right panel of Fig.~\ref{fig1}. It is evident that $\sigma^C_{yx}(\mathbf{B}=0)$  is maximum while, $\sigma^T_{yx}(\mathbf{B}=0)$ vanishes at $\theta_s=\pi/2$ and hence a pure CPHE is feasible.

We now discuss the finite $\mathbf{B}$ effect on a pure CPHE of chiral-tilted IWSM. We show the density plot $\sigma^C_{yx}(\theta_s=\pi/2)$ and $\sigma^T_{yx}(\theta_s=\pi/2)$ with $B$ and $\theta_b$ respectively in the left and right panel of Fig.(\ref{fig1-1}). The expressions of $\sigma^C_{yx}(\theta_s=\pi/2)$ and $\sigma^T_{yx}(\theta_s=\pi/2)$ are follows from Eq.~\ref{r6}: $\sigma^C_{yx}(\theta_s=\pi/2)=C_1 B B_5\cos\theta_b+C_2 B_5$ and $\sigma^T_{yx}(\theta_s=\pi/2)=C_3 B^2 \sin2\theta_b+C_4 B \sin\theta_b$ , where $C_i$'s are constants independent of $B$ and $B_5$. Therefore, a pure CPHE occurs at $\theta_b=n\pi$, shown in Fig.(\ref{fig1-1}). However, the expressions of $\sigma^C_{yx}(\theta_s=0)$ is given: $\sigma^C_{yx}(\theta_s=0)=C_3 B B_5\sin\theta_b$ and $\sigma^T_{yx}(\theta_s=0)=\sigma^T_{yx}(\theta_s=\pi/2)$. Hence, different from non-tilted IWSM, a pure CPHE is absent for $\theta_s=0$ and only emerge for $\theta_s=\pi/2$. %We leave the discussion on CLMC for the Supplemental Material~\cite{suppli1}.

The possibility of a pure CPHE, even in the absence of real magnetic fields, looks very promising. The CPHE results in the separation and accumulation of left and right chiral fermions on opposite surfaces of WSM. The chirality population difference at one surface is given as: $\delta n=\tau\sigma^{C}_{yx} E/e$. From Ref.\cite{Hosur-PRB}, we take an electric field $|E|=10$V/mm, $|B|$ and $|B_5|=1$T and assuming $\tau\sim 1 ps$, we find chirality population difference to be $\sim 450$/$\mu m^2$ and $\sim 1500$/$\mu m^2$ for $\gamma_x=0$ and $\gamma_x=0.2$, respectively. This chirality-polarization leads to an unequal optical activity on the opposite surfaces. This would manifest in a difference of absorption of left and right handed circularly polarised light at the two surfaces. Consequently, the optical activity can be detected via circular dichroism experiment\cite{Hosur-PRB}.

%\bigskip
%\begin{center}
%\begin{figure}
%\includegraphics[width=2.in]{pics.pdf}
%\caption{The chirality accumulation on the two opposite surfaces of WSM due to CPHE. The chirality imblance is detectable via unequal absorbance of left and right handed polarised light.}
%\label{chirality-acc}
%\end{figure}
%\end{center}
 
\textbf{Conclusion:}
To summarize, using semiclassical Boltzmann formalism, we establish that the strain itself can generate PHE in IWSMs without invoking the chiral anomaly. Thus the total conductivities (both PHC and LMC) get enhanced in the presence of the real electromagnetic field. Moreover, the novel strain-induced effects in IWSMs manifests itself by producing a finite CDC which is tunable by angular variation of the real magnetic field. As a consequence, a pure CPHE is realizable in IWSMs and occurs in the presence of both applied magnetic field and pseudo-magnetic field. It is worth noting that CPHE can be found in the absence of \textbf{B} in an IWSM with chiral-tilted Weyl nodes. We discuss its possible experimental verifications and provide explanations for its existence. Our general analysis provides an alternate route to explore PHE in WSMs and utilize the chiral degrees of freedom for real applications.

%We show that the total PHC gets enhanced due to chiral pseudomagnetic field. In a IWSM, applied strain induces a new anomaly $\propto \mathbf{E}\cdot\mathbf{B_5}$ by generating pseudomagnetic field ($\mathbf{B_5}$), which is different from the proper chiral anomaly ($\propto \mathbf{E}\cdot\mathbf{B}$). This novel anomaly creates an apparent charge-density non-conservation in the system. As a consequence, there is a pumping of charges between the bulk and the edge of the system which causes a finite PHE that can be detected easily. We provide analytical and numerical calculations to suggest some definite signatures of total conductivities for experimental verification. 

%We further demonstrate that a pure CPHE is possible in IWSMs. In a non-tilted as well as a tilted IWSM, CPHE occurs in presence of both applied magnetic field and pseudomagnetic field. Moreover, in an IWSM with chiral tilt, CPHE can be found in the absence of $\mathbf{B}$. We discuss its possible experimental verifications and provide explanations for its existence. Our study opens up another avenue to utilize the chiral degrees of freedom of a WSM for real applications.

%\section{ACKNOWLEDGMENTS}
\textbf{Acknowledgments:-}
 S.G. and D.S. thank Sumanta Tewari for helpful discussions. S.G. acknowledges MHRD, India for research fellowship and the computing facility from DST(India) fund for S and T infrastructure (phase-II) in the Department of Physics, IIT Kharagpur.

\end{document}

\title{Supplementary Material: Chirality-dependent planar Hall effect in inhomogeneous Weyl semimetals}
\author{Suvendu Ghosh$^1$}\email{suvenduphys@gmail.com} \author{Debabrata Sinha$^2$, Snehasish Nandy$^{1,3}$, A Taraphder$^{1,2}$}
\affiliation{$^1$ Department of Physics, Indian Institute of Technology,
Kharagpur-721302, India \\
$^2$ Center for Theoretical Studies, Indian Institute of Technology,
Kharagpur-721302, India\\
$^3$ Department of Physics, University of Virginia, Charlottesville, VA 22904, USA}
%\date{\today}
\maketitle

%\begin{widetext}
%\textcolor{blue}{
%\tableofcontents
%}
%\end{widetext}

\section{Derivations of analytical expressions for zero-temperature magnetoconductivities}
 We start with the momentum space Hamiltonian for a linearized tilted Weyl node expressed as 
\begin{eqnarray}
H_{\chi}=\hbar v_F(\chi \mathbf{k}\cdot \mathbf{\sigma}+\mathbf{\gamma}_{\chi}\cdot \mathbf{k}\sigma_0)-\mu,
\label{tilt-hamil}
\end{eqnarray}
where $v_F$ is the Fermi velocity, $\chi=\pm1$ is the chirality associated with the Weyl node, $\sigma_{i}$ represents the Pauli matrices, $\sigma_0$ is the identity matrix, and $\mu$ is the chemical potential. We choose the tilt along the $k_x$-direction.

The Boltzmann kinetic equation used here to understand the transport properties of electrons:
\begin{eqnarray}
(\frac{\partial}{\partial t}+\dot{\mathbf{r}}_\chi \cdot \mathbf{\nabla}_{\mathbf{r}}+\dot{\mathbf{k}}_\chi \cdot \mathbf{\nabla}_{\mathbf{k}})f^{\chi}_{\mathbf{r},\mathbf{k},t}=I_{coll}[f^{\chi}_{\mathbf{r},\mathbf{k},t}],
\label{steady-bol-trans}
\end{eqnarray}ms
where on the right hand side $I_{coll}[f^{\chi}_{\mathbf{r},\mathbf{k},t}]$ is the collision integral, which, within relaxation time approximation, takes the analytical form as: $I_{coll}[f^{\chi}_{\mathbf{r},\mathbf{k},t}]=(f_{eq}-f^{\chi}_k)/\tau (\mathbf{k})$ with $\tau(\mathbf{k})$ is the relaxation time. $f_\mathbf{k}^{\chi}$ is the electron distribution function, $\mathbf{k}_{\chi}$ is the quasimomentum and $f_{eq}$ is the equilibrium Fermi-Dirac distribution function that describes electron distribution in the absence of any external field. Here, we ignore the momentum dependence of the relaxation time for simplicity. Now in the presence of the effective magnetic field $\mathbf{B}^{\chi}=(\mathbf{B}+\chi \mathbf{B_5})$ and electric field $\mathbf{E}$, the semiclassical equations of motion are calculated as:
\begin{eqnarray}
\label{Eq_em_1}
&&\dot{\mathbf{r}}_{\chi}=D^{\chi}_{\mathbf{k}}[\mathbf{v}_{\mathbf{k}}+e(\mathbf{E}\times \mathbf{\Omega}^{\chi}_{\mathbf{k}})+e(\mathbf{v}_{\mathbf{k}}\cdot \mathbf{\Omega}^{\chi}_{\mathbf{k}})\mathbf{B}_{\chi}] \nonumber \\
&&\dot{\mathbf{k}}_{\chi}=D^{\chi}_{\mathbf{k}}[e\mathbf{E}+e(\mathbf{v}_{\mathbf{k}}\times \mathbf{B}_{\chi})+e^2(\mathbf{E}\cdot\mathbf{B}_{\chi})\mathbf{\Omega}^{\chi}_{\mathbf{k}}],
\label{Eq_em_2}
\end{eqnarray}
where $\mathbf{v}_{\mathbf{k}}=\frac{\partial\epsilon_\mathbf{k}}{\partial\mathbf{k}_{\chi}}$ is the group velocity and $D^{\chi}_{\mathbf{k}}=[1+e(\mathbf{B}_{\chi}\cdot\mathbf{\Omega}_{\mathbf{k}}^{\chi})]^{-1}$ depicts the modification of the phase space volume in the simultaneous presence of magnetic field and Berry curvature $\mathbf{\Omega}_{\mathbf{k}}^{\chi}$. \textcolor{black}{The charge and current density for a single chirality is given by,
\begin{eqnarray}
\label{charge-den}
\rho_{\chi}&=&e\int \frac{d^3\mathbf{k}}{(2\pi)^3}{D^{\chi}_{\mathbf{k}}}^{-1}f^\chi_\mathbf{k}\\
\mathbf{j}_{\chi}&=&e\int \frac{d^3\mathbf{k}}{(2\pi)^3}{D^{\chi}_{\mathbf{k}}}^{-1}\dot{\mathbf{r}}_{\chi}f^\chi_\mathbf{k}
\label{current-den}
\end{eqnarray}}

To calculate zero-temperature magnetoconductivities, $(\frac{\partial{f_{eq}}}{\partial{\epsilon_{\mathbf{k}}}})=-\delta(\epsilon_{\mathbf{k}}-\mu)$ because $f_{eq}=\Theta(\epsilon_{\mathbf{k}} -\mu)$ where $\Theta$ is the Heaviside step function. We have supposed that the Fermi level lies above the Weyl node, namely $\mu>0$ and the band above the node is indexed as a positive one. Consequently, we'll use only the positive sign in the dispersion, $\epsilon_{\mathbf{k}} = v_F[\gamma_x k_x \pm k]$, where $k=\sqrt{k_x^2+k_y^2+k_z^2}$. Consequently, for upper band, the velocity vector is calculated as $\mathbf{v_k}=v_F[\gamma_x+\frac{k_x}{k},\frac{k_y}{k},\frac{k_z}{k}]$ and the Berry curvature is $\Omega_k^{\chi} = \chi \frac{\mathbf{k}}{2k^3}$.\\
Because in presence of external magnetic field as well as static strain applied to the system, chiral charges feel different effective fields given by $\mathbf{B_{\chi}}=\mathbf{B}+\chi\mathbf{B_5}$, according to our setup discussed in the main text, $B_x=B\cos\theta_b+\chi B_5\cos\theta_s$ and $B_y=B\sin\theta_b+\chi B_5\sin\theta_s$.

We now define the total conductivity $\sigma^T_{ij}$ and chirality dependent conductivity $\sigma^C_{ij}$ as follows:
\begin{eqnarray}
\label{def_T}
\sigma^{T}_{ij}=\sigma^{\chi=+1}_{ij}+\sigma^{\chi=-1}_{ij}\\
\sigma^{C}_{ij}=\sigma^{\chi=+1}_{ij}-\sigma^{\chi=-1}_{ij}
\label{def_C}
\end{eqnarray}

\textcolor{black}{\subsection{Local charge nonconservation and consistent currents}
Using Eqs.(\ref{steady-bol-trans},\ref{charge-den},\ref{current-den}) and together with Maxwell's equation, one can derive the continuity equations: \begin{eqnarray}
\label{scon-1}
\partial_t\rho_5+\mathbf{\nabla}\cdot\mathbf{j_5}&=&\frac{e^2}{2\pi^2\hbar^2}(\mathbf{E}\cdot\mathbf{B}+\mathbf{E_5}\cdot\mathbf{B_5})\\
\partial_t\rho+\mathbf{\nabla}\cdot\mathbf{j}&=&\frac{e^2}{2\pi^2\hbar^2}(\mathbf{E}\cdot\mathbf{B_5}+\mathbf{E_5}\cdot\mathbf{B}),
\label{scon-2}
\end{eqnarray}
The nonconservation of chiral charge in Eq.(\ref{scon-1}) can be thought of as the chiral charge pumping between two opposite chiral Weyl nodes. On the other hand, the local non-conservation of electric charge in Eq.(\ref{scon-2}) implies that, the chrage can be created out of nothing. Therefore, to get the correct physical picture, one can define the consistent currents,  which satisfy the local charge conservation, using topological {\it Chern-Simons contributions}. Component-wise the topological charge and current densities are given by \cite{Gorbar-PRL17,Gorbar-PRB17,Gorbar-PRB18},
\begin{eqnarray}
\rho_{CS}&=&\frac{e^3}{2\pi^2\hbar^2}(\mathbf{b}\cdot \mathbf{B})\\
\mathbf{j}_{CS}&=&\frac{e^3}{2\pi^2\hbar^2}b_0\mathbf{B}-\frac{e^3}{2\pi^2\hbar^2}(\mathbf{b}\times \mathbf{E})
\label{CS-current}
\end{eqnarray}
It is now straighforward to check that the consisten current $\tilde{\mathbf{j}}^{\nu}=(\rho+\rho_{CS},\mathbf{j}+\mathbf{j}_{CS})$ is nonanomalous i.e., $\partial_{\nu}\tilde{\mathbf{j}}^{\nu}=0$. The Chern-Simons contributions $\rho_{CS}$ and $\mathbf{j}_{CS}$ have profound significance in consistent chiral kinetic theory. The first term in $\mathbf{j}_{CS}$ describes the strain-induced CME current\cite{Land-ACT-16,Cortijo_17}. However, this term is zero for a TRS broken but inversion symmetric Weyl semimetal since the energy separation between the Weyl nodes is $b_0=0$. The second term in  $\mathbf{j}_{CS}$ describes the strain-induced anomalous Hall effect in Weyl semimetal\cite{Shen-PRB17,Burkov-PRL11,Tewari-PRB13}. Contrary to the the planar Hall, the anomalous Hall conductivity has antisymmetric property (i.e., $\sigma_{yx}=-\sigma_{xy}$) and occurs in absence of $\mathbf{B}$. Therefore, the way to separate the anomalous Hall contribution from the planar Hall conductivity, is to measure the total Hall conductivity for $\mathbf{B}=0$ and $\mathbf{B}\neq 0$ and then subtracting the value for $\mathbf{B}=0$. Anyway, since the CME current and anomalous Hall current are very different from the planar Hall current, it is evident that the topological current in Eq.(\ref{CS-current}) does not contribute to both the planar Hall and longitudinal electrical conductivity in the planar setup where electromagnetic and pseudo-electromagnetic fields $E$, $B$ and $B_5$ as well as planar Hall voltage are all lie in a plane (here, we choose xy-plane).}

\subsection{Derivations of expressions for the planar Hall conductivity}
To get the analytical expression for planar Hall conductivity in an IWSM, we need to evaluate the integrals in Eq.(7) in the main text. This has been done by using the spherical polar coordinate system to write $\mathbf{k}=k[\sin\theta \cos\phi, \sin\theta \sin\phi, \cos\theta]$ and the volume element $d^3k$ as $k^2 dk \sin\theta d\theta d\phi$, where $k$ ranges from $0$ to $\infty$, $\theta$ from $0$ to $\pi$ and $\phi$ from $0$ to $2\pi$.
As the last term in Eq.(7) in the main text contains quadratic terms in $B$, we consider the expansion of $D^{\chi}_{\mathbf{k}}=[1+e(\mathbf{B}.\mathbf{\Omega_k^{\chi}})]^{-1}$ only upto the quadratic terms in $B$. Therefore, retaining only the contributions due to the quadratic terms in $B$, we write
%\begin{widetext}
\begin{eqnarray}
\sigma^{\chi}_{yx}&&=\frac{e^2\tau}{(2\pi)^3}\int d^3k \delta(\epsilon_k-\mu)[v_x v_y+ev_x B_y(\mathbf{v}_\mathbf{k}.\mathbf{\Omega_k^{\chi}})\nonumber\\
&&+ev_y B_x(\mathbf{v}_\mathbf{k}.\mathbf{\Omega_k^{\chi}})+e^2 B_xB_y(\mathbf{v}_\mathbf{k}.\mathbf{\Omega_k^{\chi}})^2]\nonumber\\
&&-\frac{e^3\tau}{(2\pi)^3}\int d^3k \delta(\epsilon_k-\mu)(\mathbf{B}.\mathbf{\Omega_k^{\chi}})[v_x v_y+ev_x B_y(\mathbf{v}_\mathbf{k}.\mathbf{\Omega_k^{\chi}})\nonumber\\
&&+ev_y B_x(\mathbf{v}_\mathbf{k}.\mathbf{\Omega_k^{\chi}})]+\frac{e^4\tau}{(2\pi)^3}\int d^3k \delta(\epsilon_k-\mu)[v_x v_y](\mathbf{B}.\mathbf{\Omega_k^{\chi}})^2\nonumber\\
\label{appex_eqn_1}
\end{eqnarray}
%\end{widetext}

Now, using the spherical polar coordinate system and getting the integration over $k$ done, we have
\begin{widetext}
\begin{eqnarray}
\sigma^{\chi}_{yx}&&=\frac{e^2v_F\tau}{(2\pi)^3}\int\limits_0^{2\pi}d\phi\int\limits_0^{\pi}d\theta[\frac{\gamma_x\mu^2}{v^2_F}\frac{\sin^2\theta\sin\phi}{(\gamma_x\sin\theta\cos\phi+1)^3}+\frac{\mu^2}{v^2_F}\frac{\sin^3\theta\cos\phi\sin\phi}{(\gamma_x\sin\theta\cos\phi+1)^3}+\frac{e\chi B_y\gamma_x^2}{2}\frac{\sin^2\theta\cos\phi}{(\gamma_x\sin\theta\cos\phi+1)}\nonumber\\
&&+\frac{e\chi B_y\gamma_x}{2}\frac{\sin\theta}{(\gamma_x\sin\theta\cos\phi+1)}+\frac{e\chi B_y\gamma_x}{2}\frac{\sin^3\theta\cos^2\phi}{(\gamma_x\sin\theta\cos\phi+1)}+\frac{e\chi B_y}{2}\frac{\sin^2\theta\cos\phi}{(\gamma_x\sin\theta\cos\phi+1)}+\frac{e\chi B_y\gamma_x}{2}\frac{\sin^3\theta\cos\phi\sin\phi}{(\gamma_x\sin\theta\cos\phi+1)}\nonumber\\
&&+\frac{e\chi B_x}{2}\frac{\sin^2\theta\sin\phi}{(\gamma_x\sin\theta\cos\phi+1)}+\frac{e^2B_xB_y\gamma_x^2v^2_F}{4\mu^2}(\gamma_x\sin\theta\cos\phi+1)\sin^3\theta\cos^2\phi+\frac{e^2B_xB_y\gamma_xv^2_F}{2\mu^2}(\gamma_x\sin\theta\cos\phi+1)\sin^2\theta\nonumber\\
&&\cos\phi+\frac{e^2B_xB_yv^2_F}{4\mu^2}(\gamma_x\sin\theta\cos\phi+1)\sin\theta]-\frac{e^3v_F\tau\chi}{2(2\pi)^3}\int\limits_0^{2\pi}d\phi\int\limits_0^{\pi}d\theta[\gamma_xB_x\frac{\sin^3\theta\cos\phi\sin\phi}{(\gamma_x\sin\theta\cos\phi+1)}+B_x\frac{\sin^4\theta\cos^2\phi\sin\phi}{(\gamma_x\sin\theta\cos\phi+1)}\nonumber\\
&&+\gamma_xB_y\frac{\sin^3\theta\sin^2\phi}{(\gamma_x\sin\theta\cos\phi+1)}+B_y\frac{\sin^4\theta\cos\phi\sin^2\phi}{(\gamma_x\sin\theta\cos\phi+1)}]-\frac{e^4v^3_F\tau}{4(2\pi)^3\mu^2}\int\limits_0^{2\pi}d\phi\int\limits_0^{\pi}d\theta[B_xB_y\gamma_x^2(\gamma_x\sin\theta\cos\phi+1)\sin^3\theta\cos^2\phi\nonumber\\
&&+B_xB_y\gamma_x(\gamma_x\sin\theta\cos\phi+1)\sin^2\theta\cos\phi+B_xB_y\gamma_x(\gamma_x\sin\theta\cos\phi+1)\sin^4\theta\cos^3\phi+B_xB_y(\gamma_x\sin\theta\cos\phi+1)\sin^3\theta\nonumber\\
&&\cos^2\phi+B_y^2\gamma_x^2(\gamma_x\sin\theta\cos\phi+1)\sin^3\theta\cos\phi\sin\phi+B_y^2\gamma_x(\gamma_x\sin\theta\cos\phi+1)\sin^2\theta\sin\phi+B_y^2\gamma_x(\gamma_x\sin\theta\cos\phi+1)\nonumber\\
&&\sin^4\theta\cos^2\phi\sin\phi+B_y^2(\gamma_x\sin\theta\cos\phi+1)\sin^3\theta\cos\phi\sin\phi+B_x^2\gamma_x(\gamma_x\sin\theta\cos\phi+1)\sin^4\theta\cos^2\phi\sin\phi+B_x^2(\gamma_x\sin\theta\nonumber\\
&&\cos\phi+1)\sin^3\theta\cos\phi\sin\phi+B_xB_y\gamma_x(\gamma_x\sin\theta\cos\phi+1)\sin^4\theta\cos\phi\sin^2\phi+B_xB_y(\gamma_x\sin\theta\cos\phi+1)\sin^3\theta\sin^2\phi]\nonumber\\
&&+\frac{e^4v^3_F\tau}{4(2\pi)^3\mu^2}\int\limits_0^{2\pi}d\phi\int\limits_0^{\pi}d\theta[B^2_x\gamma_x(\gamma_x\sin\theta\cos\phi+1)\sin^4\theta\cos^2\phi\sin\phi+B_x^2(\gamma_x\sin\theta\cos\phi+1)\sin^5\theta\cos^3\phi\sin\phi+2B_xB_y\nonumber\\
&&\gamma_x(\gamma_x\sin\theta\cos\phi+1)\sin^4\theta\cos\phi\sin^2\phi+2B_xB_y(\gamma_x\sin\theta\cos\phi+1)\sin^5\theta\cos^2\phi\sin^2\phi+B_y^2\gamma_x(\gamma_x\sin\theta\cos\phi+1)\nonumber\\
&&\sin^4\theta\sin^3\phi+B_y^2(\gamma_x\sin\theta\cos\phi+1)\sin^5\theta\cos\phi\sin^3\phi]
\label{appex_eqn_2}
\end{eqnarray}
\end{widetext}
$\theta$- and $\phi$-integrals can now readily be evaluated for an isotropic type-I IWSM, i.e., for $\gamma_x=0$. After getting those integrals done, adequate simplification makes us obtain the planar Hall conductivity for a Weyl node as 
\begin{eqnarray}
\sigma^{\chi}_{yx}&&=\frac{7e^4v^3_F\tau}{120\pi^2\mu^2}[B^2\sin2\theta_b+B_5^2\sin2\theta_s\nonumber\\&&+\chi BB_5\sin(\theta_b+\theta_s)]\nonumber\\
\label{appex_eqn_3}
\end{eqnarray}

Again for a tilted type-I Weyl node, as $|\gamma_x|<1$, expanding $(\gamma_x\sin\theta \cos\phi+1)^{-n}$ (here $n=1$ and $3$ only) upto the terms which contain $\gamma_x^2$ and neglecting the terms containing the higher orders of $\gamma_x$, we reach the result, planar Hall conductivity, as
\begin{eqnarray}
\sigma^{\chi}_{yx}&&=\frac{e^4v^3_F\tau}{80\pi^2\mu^2}(\frac{7}{3}+\gamma_x^2)[B^2\sin2\theta_b+B_5^2\sin2\theta_s\nonumber\\
&&+\chi BB_5\sin(\theta_b+\theta_s)]+\chi\frac{e^3v_F\tau}{30\pi^2}(\frac{11\gamma_x}{2}+\gamma_x^3)\nonumber\\
&&B\sin\theta_b+\frac{e^3v_F\tau}{30\pi^2}(\frac{11\gamma_x}{2}+\gamma_x^3)B_5\sin\theta_s
\label{appex_eqn_4}
\end{eqnarray}

\subsection{Derivations of expressions for the longitudinal magneto-conductivity}

To get the analytical expression for longitudinal magneto-conductivity in an IWSM, we need to evaluate the integrals in Eq.(6) in the main text. This has been done again by using the spherical polar coordinate system as we've done in the previous section. As the last term in Eq.(6) in the main text contains quadratic terms in $B$, we consider the expansion of $D^{\chi}_{\mathbf{k}}=D(\mathbf{B},\mathbf{\Omega_k^{\chi}})=[1+e(\mathbf{B}.\mathbf{\Omega_k^{\chi}})]^{-1}$ only upto the quadratic terms in $B$. Therefore, retaining only the contributions due to the quadratic terms in $B$, we write
\begin{eqnarray}
\sigma^{\chi}_{xx}&&=\frac{e^2\tau}{(2\pi)^3}\int d^3k \delta(\epsilon_k-\mu)[v_x^2+2ev_x(\mathbf{v}_\mathbf{k}.\mathbf{\Omega_k^{\chi}})B_x\nonumber\\&&+e^2(\mathbf{v}_\mathbf{k}.\mathbf{\Omega_k^{\chi}})^2B_x^2]-\frac{e^3\tau}{(2\pi)^3}\int d^3k \delta(\epsilon_k-\mu)(\mathbf{B}\cdot\mathbf{\Omega_k^{\chi}})\nonumber\\&&[v_x^2+2ev_x(\mathbf{v}_\mathbf{k}.\mathbf{\Omega_k^{\chi}})B_x]+\frac{e^4\tau}{(2\pi)^3}\int d^3k \delta(\epsilon_k-\mu)\nonumber\\&&v_x^2(\mathbf{B}\cdot\mathbf{\Omega_k^{\chi}})^2
\label{appex_eqn_5}
\end{eqnarray}
Now, using the spherical polar coordinate system and getting the integration over $k$ done, we have
\begin{widetext}
\begin{eqnarray}
\sigma^{\chi}_{xx}&&=\frac{e^2v_F\tau}{(2\pi)^3}\int\limits_0^{2\pi}d\phi\int\limits_0^{\pi}d\theta[\frac{\gamma_x^2\mu^2}{v_F^2}\frac{\sin\theta}{(\gamma_x\sin\theta\cos\phi+1)^3}+\frac{2\gamma_x\mu^2}{v_F^2}\frac{\sin^2\theta\cos\phi}{(\gamma_x\sin\theta\cos\phi+1)^3}+\frac{\mu^2}{v_F^2}\frac{\sin^3\theta\cos^2\phi}{(\gamma_x\sin\theta\cos\phi+1)^3}\nonumber\\
&&+e\chi B_x\gamma_x^2\frac{\sin^2\theta\cos\phi}{(\gamma_x\sin\theta\cos\phi+1)}+e\chi B_x\gamma_x\frac{\sin\theta}{(\gamma_x\sin\theta\cos\phi+1)}+e\chi B_x\gamma_x\frac{\sin^3\theta\cos^2\phi}{(\gamma_x\sin\theta\cos\phi+1)}+e\chi B_x\frac{\sin^2\theta\cos\phi}{(\gamma_x\sin\theta\cos\phi+1)}\nonumber\\
&&+\frac{e^2B_x^2\gamma_x^2v_F^2}{4\mu^2}(\gamma_x\sin\theta\cos\phi+1)\sin^3\theta\cos^2\phi+\frac{e^2B_x^2\gamma_xv_F^2}{2\mu^2}(\gamma_x\sin\theta\cos\phi+1)\sin^2\theta\cos\phi\nonumber\\
&&+\frac{e^2B_x^2v_F^2}{4\mu^2}(\gamma_x\sin\theta\cos\phi+1)\sin\theta]-\frac{e^3v_F\tau\chi}{(16\pi)^3}\int\limits_0^{2\pi}d\phi\int\limits_0^{\pi}d\theta[\gamma_x^2B_x\frac{\sin^2\theta\cos\phi}{(\gamma_x\sin\theta\cos\phi+1)}+2\gamma_xB_x\frac{\sin^3\theta\cos^2\phi}{(\gamma_x\sin\theta\cos\phi+1)}\nonumber\\
&&+B_x\frac{\sin^4\theta\cos^3\phi}{(\gamma_x\sin\theta\cos\phi+1)}+\gamma_x^2B_y\frac{\sin^2\theta\sin\phi}{(\gamma_x\sin\theta\cos\phi+1)}+2\gamma_xB_y\frac{\sin^3\theta\cos\phi\sin\phi}{(\gamma_x\sin\theta\cos\phi+1)}+B_y\frac{\sin^4\theta\cos^2\phi\sin\phi}{(\gamma_x\sin\theta\cos\phi+1)}\nonumber\\
&&+\frac{e\chi B_x^2\gamma_x^2v_F^2}{\mu^2}(\gamma_x\sin\theta\cos\phi+1)\sin^3\theta\cos^2\phi+\frac{e\chi B_x^2\gamma_xv_F^2}{\mu^2}(\gamma_x\sin\theta\cos\phi+1)\sin^2\theta\cos\phi\nonumber\\
&&+\frac{e\chi B_x^2\gamma_xv_F^2}{\mu^2}(\gamma_x\sin\theta\cos\phi+1)\sin^4\theta\cos^3\phi+\frac{e\chi B_x^2v_F^2}{\mu^2}(\gamma_x\sin\theta\cos\phi+1)\sin^3\theta\cos^2\phi\nonumber\\
&&+\frac{e\chi B_xB_y\gamma_x^2v_F^2}{\mu^2}(\gamma_x\sin\theta\cos\phi+1)\sin^3\theta\cos\phi\sin\phi+\frac{e\chi B_xB_y\gamma_xv_F^2}{\mu^2}(\gamma_x\sin\theta\cos\phi+1)\sin^2\theta\sin\phi\nonumber\\
&&+\frac{e\chi B_xB_y\gamma_xv_F^2}{\mu^2}(\gamma_x\sin\theta\cos\phi+1)\sin^4\theta\cos^2\phi\sin\phi+\frac{e\chi B_xB_yv_F^2}{\mu^2}(\gamma_x\sin\theta\cos\phi+1)\sin^3\theta\cos\phi\sin\phi]\nonumber\\
&&+\frac{e^4v_F^3\tau}{32\pi^3\mu^2}\int\limits_0^{2\pi}d\phi\int\limits_0^{\pi}d\theta[\gamma_x^2B_x^2(\gamma_x\sin\theta\cos\phi+1)\sin^3\theta\cos^2\phi+2\gamma_xB_x^2(\gamma_x\sin\theta\cos\phi+1)\sin^4\theta\cos^3\phi\nonumber\\
&&+B_x^2(\gamma_x\sin\theta\cos\phi+1)\sin^5\theta\cos^4\phi+2\gamma_x^2B_xB_y(\gamma_x\sin\theta\cos\phi+1)\sin^3\theta\cos\phi\sin\phi\nonumber\\
&&+4\gamma_xB_xB_y(\gamma_x\sin\theta\cos\phi+1)\sin^4\theta\cos^2\phi\sin\phi+2B_xB_y(\gamma_x\sin\theta\cos\phi+1)\sin^5\theta\cos^3\phi\sin\phi\nonumber\\
&&+\gamma_x^2B_y^2(\gamma_x\sin\theta\cos\phi+1)\sin^3\theta\sin^2\phi+2\gamma_xB_y^2(\gamma_x\sin\theta\cos\phi+1)\sin^4\theta\cos\phi\sin^2\phi\nonumber\\
&&+B_y^2(\gamma_x\sin\theta\cos\phi+1)\sin^5\theta\cos^2\phi\sin^2\phi]
\label{appex_eqn_6}
\end{eqnarray}
\end{widetext}
\textcolor{black}{Though $\sigma^{\chi}_{xx}$ contains only $B_x$-terms when the phase space volume is not modified, (i.e., $D^{\chi}_{\mathbf{k}}=1$), but the expansion of $D^{\chi}_{\mathbf{k}}$ upto $B^2$-terms generates $B_y$-dependent terms in the final experession of $\sigma^{\chi}_{xx}$.} $\theta$- and $\phi$-integrals can now readily be evaluated for an isotropic type-I IWSM, i.e., for $\gamma_x=0$. After getting those integrals done, adequate simplification makes us obtain the longitudinal magneto-conductivity for a Weyl node as
%\begin{widetext}
\begin{eqnarray}
\sigma^{\chi}_{xx}&&=\frac{e^2\mu^2\tau}{6\pi^2v_F}+\frac{7e^4v_F^3\tau}{120\pi^2\mu^2}[B^2\cos^2\theta_b+B^{2}_{5}\cos^2\theta_s]\nonumber\\
&&+\frac{e^4v_F^3\tau}{120\pi^2\mu^2}[B^2+B_5^2]+\chi\frac{2e^4v_F^3\tau}{15\pi^2\mu^2}BB_5\cos\theta_b\cos\theta_s\nonumber\\
&&+\chi\frac{e^4v_F^3\tau}{60\pi^2\mu^2}BB_5\sin\theta_b\sin\theta_s
\label{appex_eqn_7}
\end{eqnarray}
%\end{widetext}

Again for a tilted type-I Weyl node, as $|\gamma_x|<1$, expanding $(\gamma_x\sin\theta \cos\phi+1)^{-n}$ (here $n=1$ and $3$ only) upto the terms which contain $\gamma_x^2$ and neglecting the terms containing the higher orders of $\gamma_x$, we reach the result, longitudinal magneto-conductivity, as
\begin{eqnarray}
\sigma^{\chi}_{xx}&&=\frac{e^2\mu^2\tau}{\pi^2v_F}(\frac{1}{6}+\gamma_x^4)+\frac{e^4v_F^3\tau}{15\pi^2\mu^2}[B^2\cos^2\theta_b+B^{2}_{5}\cos^2\theta_s]\nonumber\\
&&+\chi\frac{2e^4v_F^3\tau}{15\pi^2\mu^2}BB_5\cos\theta_b\cos\theta_s+\frac{e^4v_F^3\tau}{120\pi^2\mu^2}(1+7\gamma_x^2)\nonumber\\
&&[B^2\sin^2\theta_b+B^{2}_{5}\sin^2\theta_s+2\chi BB_5\sin\theta_b\sin\theta_s]\nonumber\\
&&+\frac{e^3v_F\tau}{12\pi^2}(\frac{23}{5}\gamma_x+\gamma_x^3)[\chi B\cos\theta_b+B_5\cos\theta_s]
\label{appex_eqn_8}
\end{eqnarray}
%We obtain the expressions of total and chiral LMC in Eqs.(9,11,14,16) of the main text using Eqs.(\ref{appex_eqn_7},\ref{appex_eqn_8}).

\section{Discussion: magneto-conductivity in a type-I IWSM}

\begin{figure}
\includegraphics[width=1.65in]{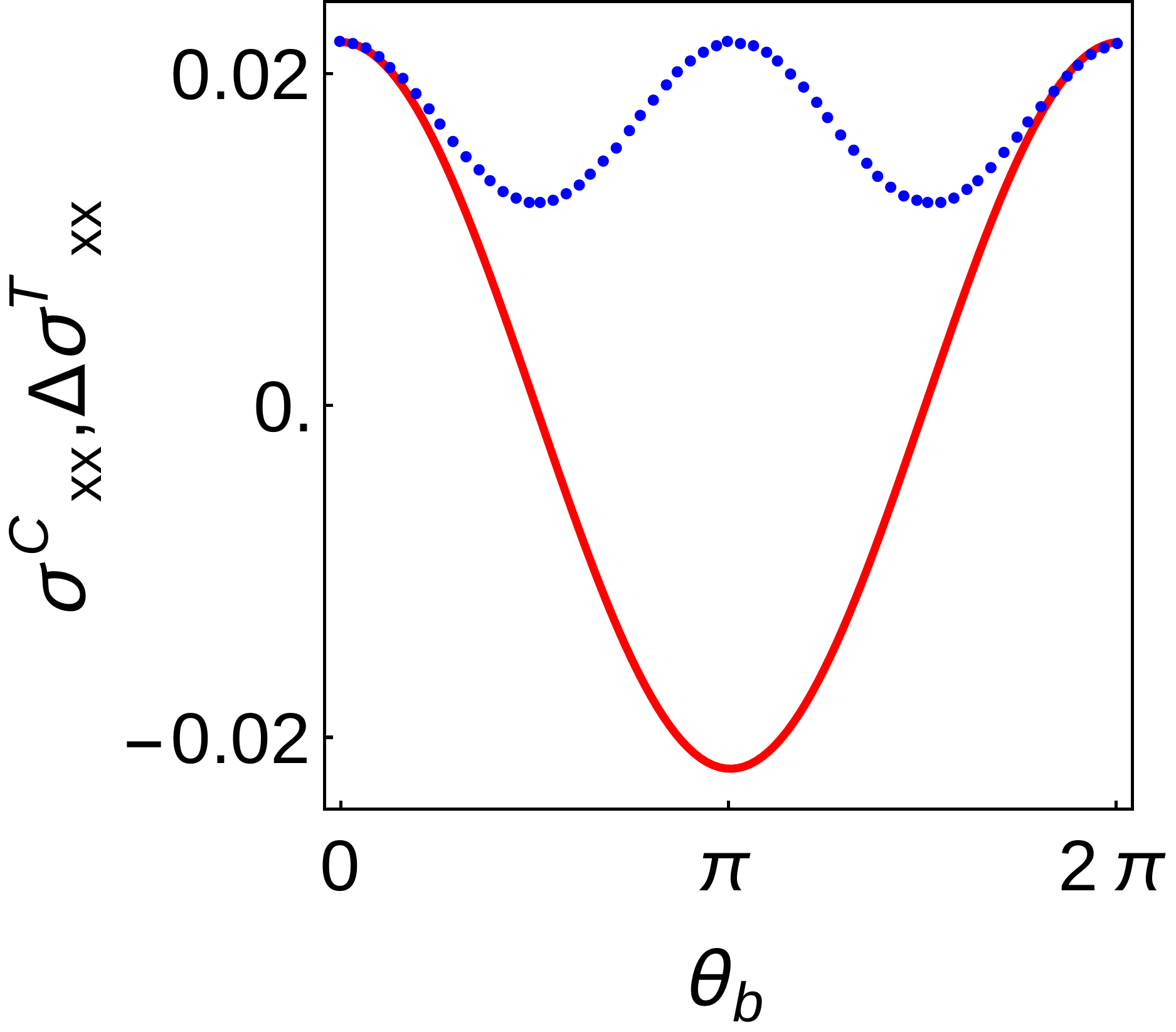}
\includegraphics[width=1.65in]{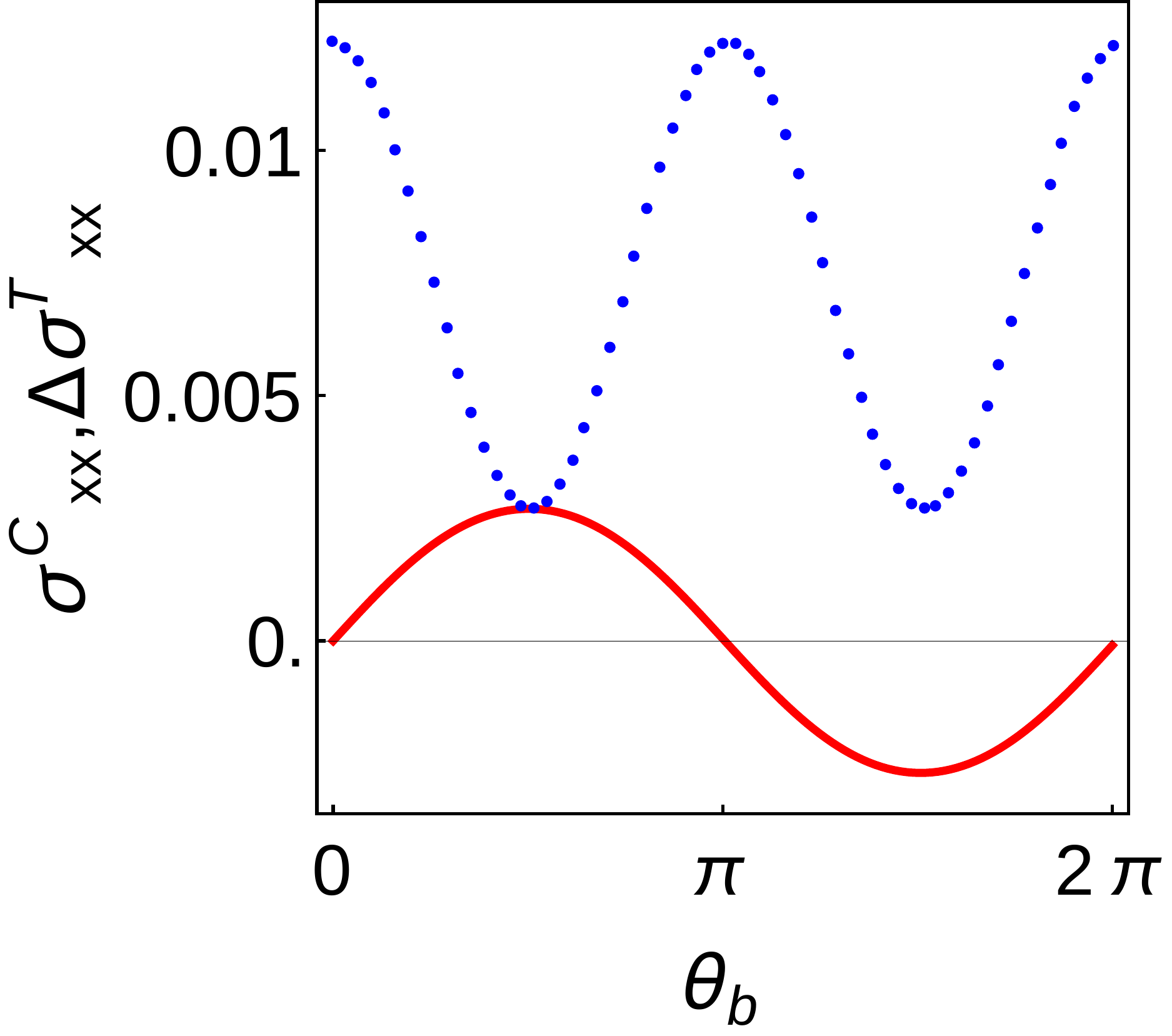}
\caption{Non-tilted IWSM. The angular ($\theta_b$) dependence of $\sigma^C_{xx}$ (red solid line), $\Delta\sigma^T_{xx}$ (blue dotted line). We fix $\theta_s=0$ and  $\theta_s=\pi/2$ in left and right panel respectively. The other parameters are $B=B_5=0.7$T, $\mu=20$meV and $\gamma_x=0$. All the conductivities are normalized by the corresponding total longitudinal conductivity without effective magnetic field $\sigma^T_{xx}(B=B_5=0)$.}
\label{long-nt}
\end{figure}

\begin{figure}
\includegraphics[width=1.65in]{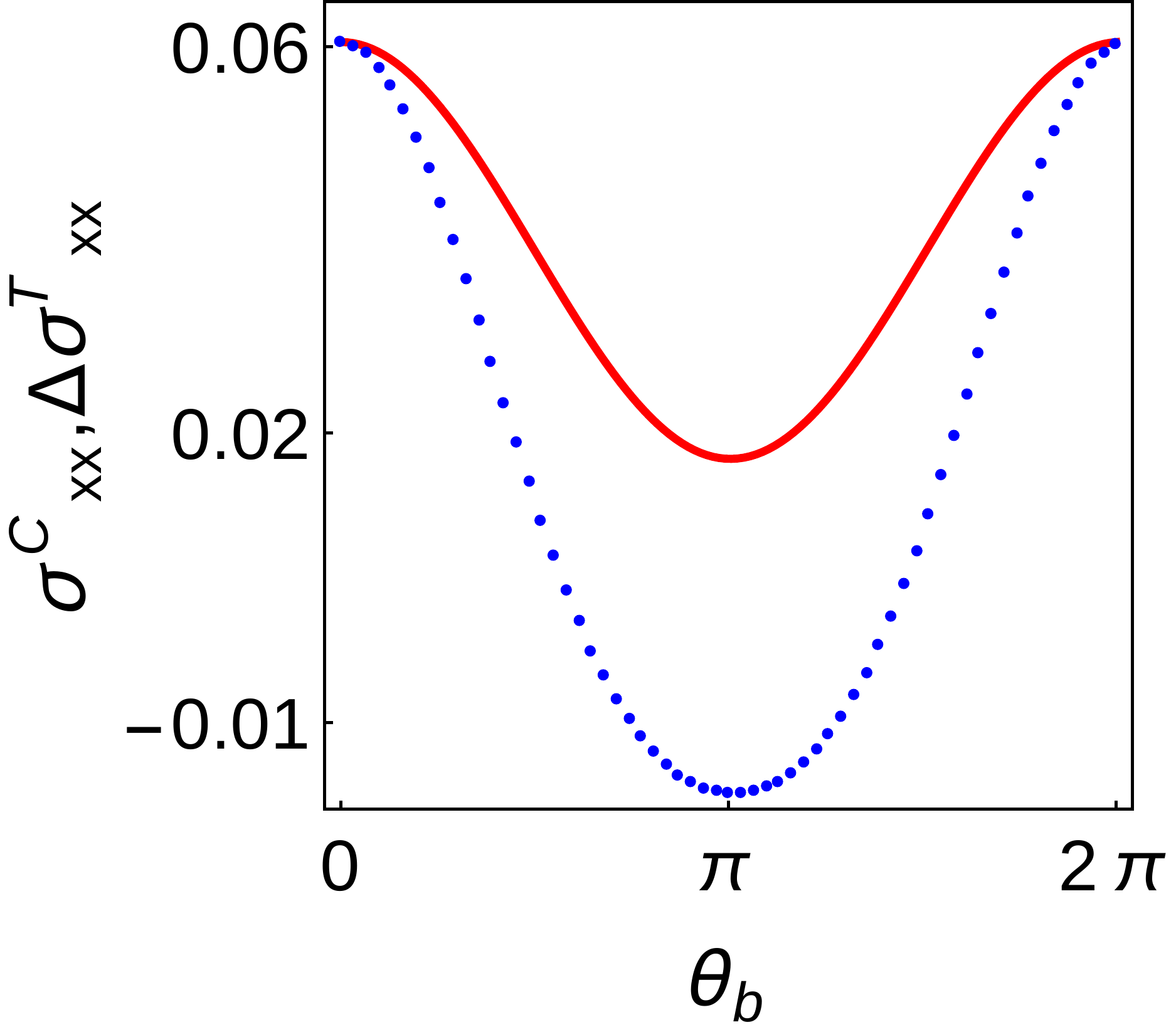}
\includegraphics[width=1.65in]{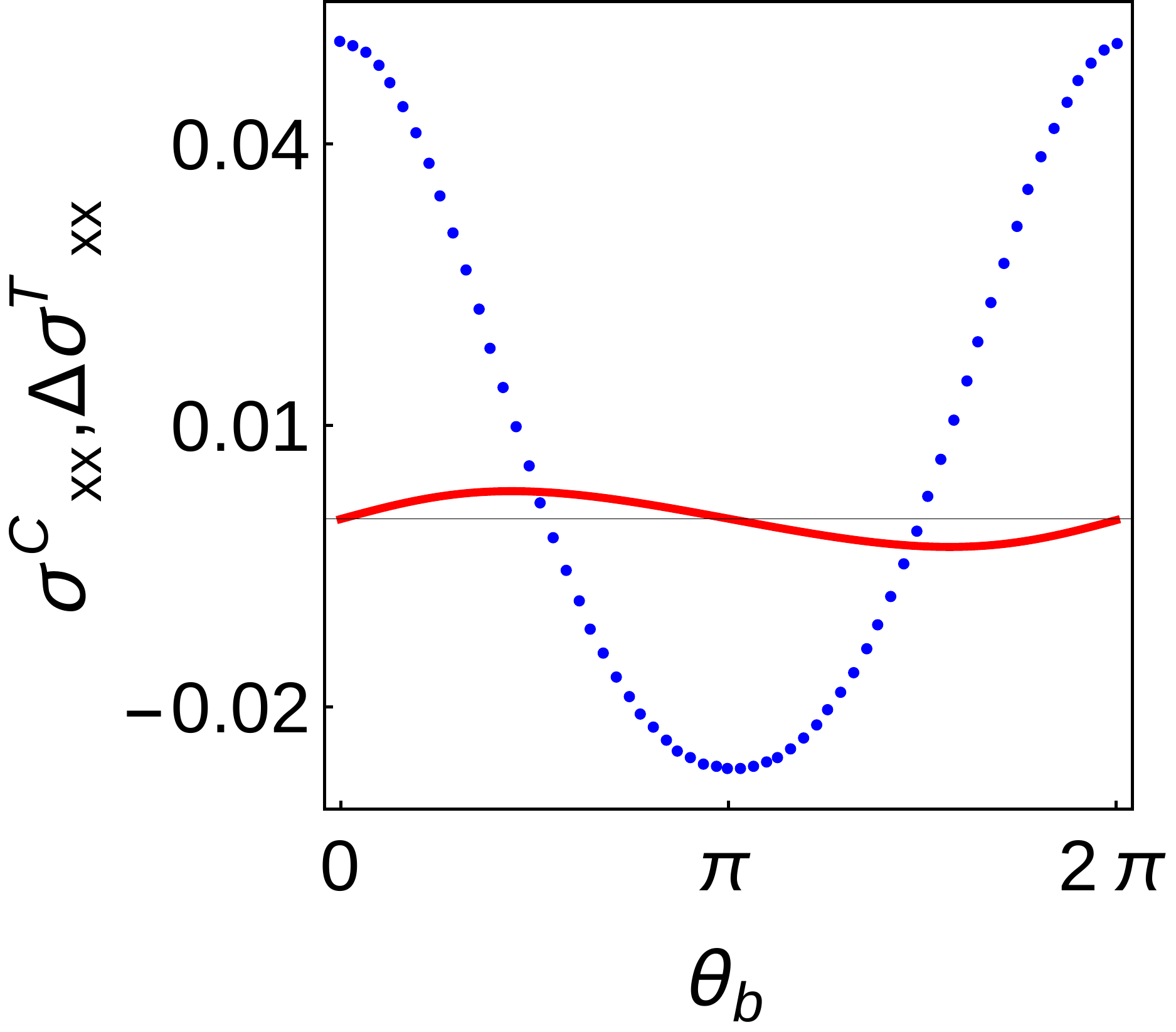}
\caption{Tilted IWSMs (chiral tilt). The angular ($\theta_b$) dependence of $\sigma^C_{xx}$ (red solid line), $\Delta\sigma^T_{xx}$ (blue dotted line). We fix $\theta_s=0$ and  $\theta_s=\pi/2$ in left and right panel respectively. The other parameters are $B=B_5=0.7$T, $\mu=20$meV and $\gamma_x=0.2$. All the conductivities are normalized by the corresponding total longitudinal conductivity without effective magnetic field $\sigma^T_{xx}(B=B_5=0)$.}
\label{long-tt}
\end{figure}

\textcolor{blue}{Non-tilted IWSM:} We have shown the angular variation ($\theta_b$) of $\sigma^C_{xx}$ and $\Delta\sigma^T_{xx}$ for a non-tilted IWSM in Fig.(\ref{long-nt}) with $\theta_s=0$ and $\theta_s=\pi/2$. $\sigma^C_{xx}$ varies as $\cos\theta_b$ and $\sin \theta_b$ respectively for these two values of $\theta_s$ whereas $\Delta\sigma^T_{xx}$ shows $\cos^2\theta_b$ variation. Thus, $\Delta\sigma^T_{xx}$ remains positive (i.e., the negative magnetoresistance) for the entire range of $\theta_b$.

\textcolor{blue}{IWSM with chiral tilt:} We have shown the angular variation ($\theta_b$) of  $\sigma^C_{xx}$ and $\Delta\sigma^T_{xx}$ for an IWSM with chiral tilt in Fig.(\ref{long-tt}). $\sigma^C_{xx}$ varies as $\cos\theta_b$ and $\sin \theta_b$ for  $\theta_s=0$ and $\theta_s=\pi/2$ respectively. However, $\Delta\sigma^T_{xx}$ is clearly seen to vary as $\cos \theta_b$. Thus, $\Delta\sigma^T_{xx}$ becomes negative with $\theta_b$.

\textcolor{blue}{IWSM with non-chiral tilt:} In case of non-chiral tilt, the expressions for total LMC and PHC are calculated as
\begin{eqnarray}
\label{r7}
\sigma_{xx}^T&=\frac{e^2\mu^2\tau}{\pi^2v_F}(\frac{1}{3}+2\gamma_x^4)+\frac{2e^4v^3_F\tau}{15\pi^2\mu^2}(B^2\cos^2\theta_b+B_5^2\cos^2\theta_s)\nonumber\\
&+\frac{e^4v^3_F\tau}{60\pi^2\mu^2}(1+7\gamma_x^2)(B^2\sin^2\theta_b+B_5^2\sin^2\theta_s)\nonumber\\
&+\frac{e^3v_F\tau}{6\pi^2}(\frac{23}{5}\gamma_x+\gamma_x^3)B_5\cos\theta_s\\
\sigma_{yx}^T&=\frac{e^4v^3_F\tau}{40\pi^2\mu^2}(\frac{7}{3}+\gamma_x^2)(B^2\sin 2\theta_b+B_5^2\sin 2\theta_s)\nonumber\\&+\frac{e^3 v_{F}\tau}{15\pi^2}(\frac{11}{2}\gamma_x+\gamma_x^3)B_5\sin\theta_s
\label{r8}
\end{eqnarray}

However, the expressions for chirality-dependent LMC and PHC are calculated as
\begin{eqnarray}
\label{r11}
\sigma_{xx}^{C}&=&\frac{4e^4v^3_F\tau}{15\pi^2\mu^2}BB_5\cos\theta_b\cos\theta_s\nonumber\\&&+\frac{e^3v_F\tau}{6\pi^2}(\frac{23}{5}\gamma_x+\gamma_x^3)B\cos\theta_b\nonumber\\
&&+\frac{4e^4v^3_F\tau}{30\pi^2\mu^2}(1+7\gamma_x^2)BB_5\sin\theta_b\sin\theta_s\\
\sigma_{yx}^{C}&=&\frac{e^4v^3_F\tau}{20\pi^2\mu^2}(\frac{7}{3}+\gamma_x^2)BB_5\sin(\theta_b+\theta_s)\nonumber\\&&+\frac{e^3v_F\tau}{15\pi^2}(\frac{11}{2}\gamma_x+\gamma_x^3)B\sin\theta_b .
\label{r12}
\end{eqnarray}

\begin{figure}
\includegraphics[width=1.65in]{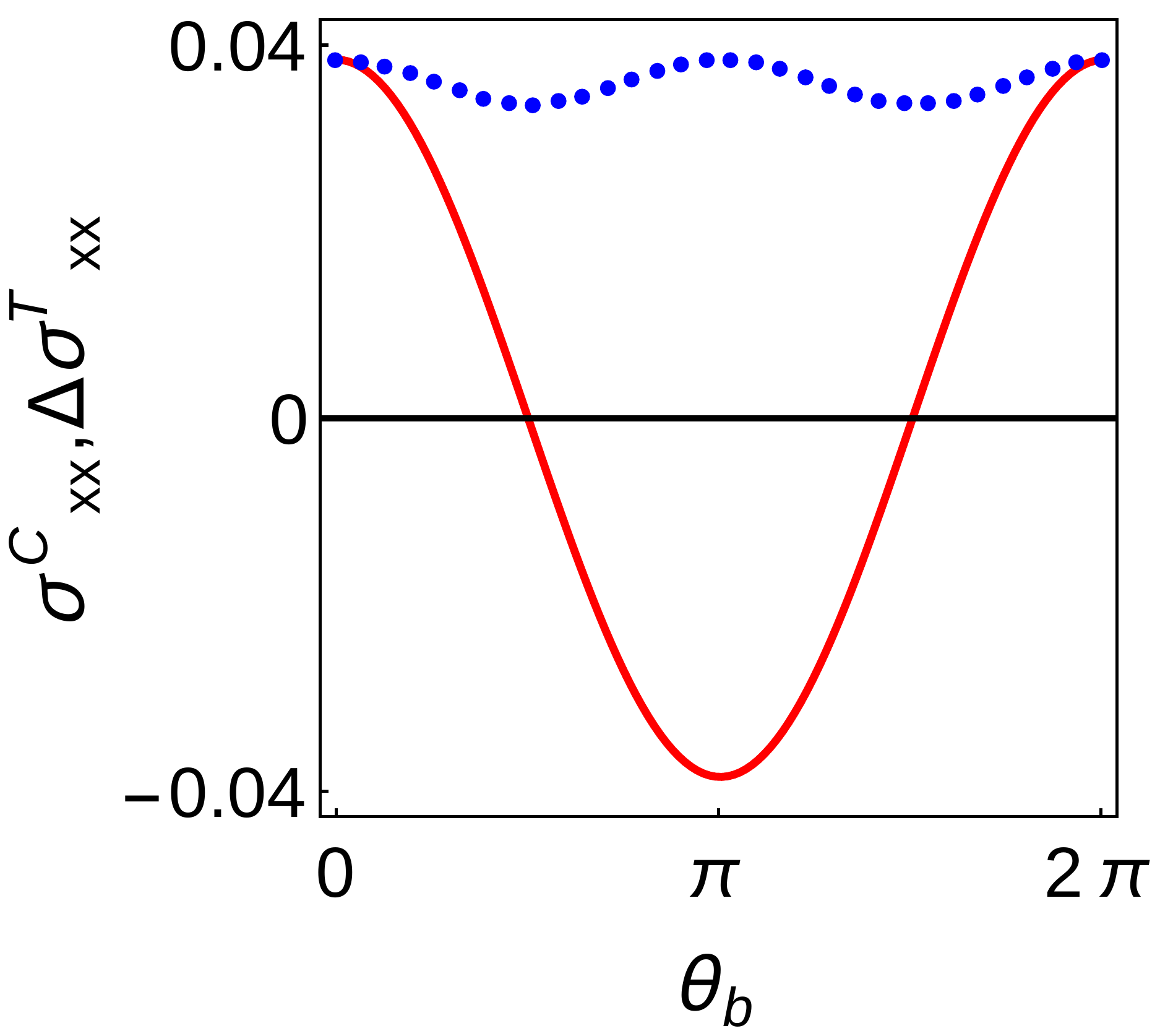}
\includegraphics[width=1.65in]{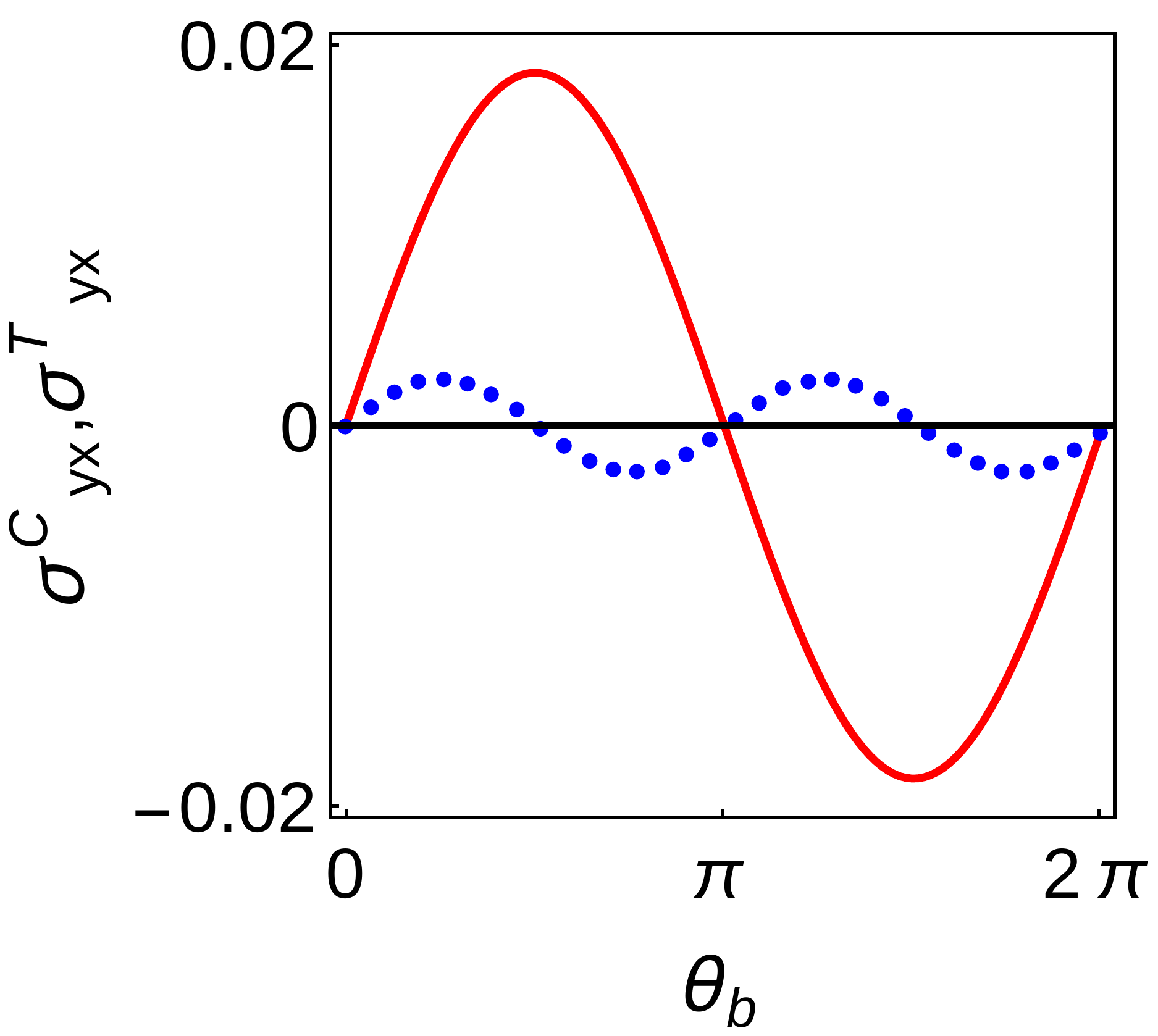}
\caption{Tilted IWSMs (non-chiral tilt). Left panel shows the angular ($\theta_b$) dependence of $\sigma^C_{xx}$ (red solid line), $\Delta\sigma^T_{xx}$ (blue dotted line). Right panel shows the angular ($\theta_b$) dependence of $\sigma^C_{yx}$ (red solid line), $\Delta\sigma^T_{yx}$ (blue dotted line). Here, tilt, $\gamma_x=0.2$, $\mu=20meV$, $B=B_5=0.5$T and $\theta_s=0$.}
\label{non-chiral}
\end{figure}

Left panel of Fig.(\ref{non-chiral}) shows the angular variation ($\theta_b$) of $\sigma^C_{xx}$ (red solid line) and $\Delta\sigma^T_{xx}$ (blue dotted line) for a type-I IWSM with non-chiral tilt.  When strain is absent ($B_5=0$), $\sigma^T_{xx}$ and $\sigma^C_{xx}$ still exist as it is evident from the Eq.(\ref{r7}) and (\ref{r11}). Our numerical calculation shows that $\sigma^C_{xx}$ varies as $\cos\theta_b$ for all values of $B$, $B_5$ and $\theta_s$ which clearly indicates that the second term in Eq.(\ref{r11}) is dominating. Eq.(\ref{r11}) also informs that $\sigma^C_{xx}$ will vanish in absence of external magnetic field with which it varies linearly. On the other hand, $\Delta\sigma^T_{xx}$ varies as $\cos^2\theta_b$ for all values of $B$, $B_5$ and $\theta_s$, as shown in the plot. It has quadratic variation with $B$ and linear variation with $B_5$. Interestingly, Eq.(\ref{r7}) reveals that $\sigma^T_{xx}$ exist even when $B$ is not applied.

Right panel of Fig.(\ref{non-chiral}) shows the angular variation ($\theta_b$) of $\Delta\sigma^T_{yx}$ (blue dotted line) and $\sigma^C_{yx}$ (red solid line) for a type-I IWSM with non-chiral tilt. When strain is absent ($B_5=0$), $\sigma^T_{yx}$ and $\sigma^C_{yx}$ still exist as it is evident from the Eq.(\ref{r8}) and (\ref{r12}). Our numerical calculation shows that $\sigma^C_{yx}$ varies as $\sin\theta_b$ for all values of $B$, $B_5$ and $\theta_s$ which clearly indicates that the last term in Eq.(\ref{r12}) is dominating. Eq.(\ref{r12}) also informs that $\sigma^C_{yx}$ will vanish in absence of external magnetic field with which it varies linearly. On the other hand, $\Delta\sigma^T_{yx}$ varies as $\sin2\theta_b$ for all values of $B$, $B_5$ and $\theta_s$, as shown in the plot. It has quadratic variation with $B$ and linear variation with $B_5$. Interestingly, Eq.(\ref{r8}) reveals that $\sigma^T_{yx}$ can exist even when $B$ is not applied.

%\begin{figure}
%\includegraphics[width=1.65in]{supp-fig1.pdf}
%\includegraphics[width=1.65in]{supp-fig2.pdf}
%\includegraphics[width=1.65in]{supp-fig6.pdf}
%\includegraphics[width=1.65in]{supp-fig3.pdf}
%\caption{Tilted IWSMs (non-chiral tilt). Left panel shows the angular ($\theta_b$) dependence of $\sigma^C_{xx}$ (red solid line), $\Delta\sigma^T_{xx}$ (blue dotted line). Right panel shows the angular ($\theta_b$) dependence of $\sigma^C_{yx}$ (red solid line), $\Delta\sigma^T_{yx}$ (blue dotted line). Here, tilt, $\gamma_x=0.2$, $\mu=20meV$, $B=B_5=0.5$T and $\theta_s=\pi/2$.}
%\label{nc-ph}
%\end{figure}